\documentclass[sigconf,screen,noacm]{acmart}

\usepackage{soul}
\usepackage{tikz}
\usepackage{subfig}
\usepackage{xspace}
\usepackage{siunitx}

\AtBeginDocument{%
  }

\setcopyright{none}
\renewcommand\footnotetextcopyrightpermission[1]{}
\graphicspath{{figs/}}
\ifx\figurename\undefined \def\figurename{Figure}\fi
\renewcommand{\figurename}{Fig.}
\newcommand{\para}[1]{\textit{\textbf{#1}} }
\renewcommand{\subparagraph}[1]{\underline{\textit{#1}} }

\newcommand{\Sect}[1]{Sec.~\ref{#1}}
\newcommand{\Fig}[1]{Fig.~\ref{#1}}
\newcommand{\Tbl}[1]{Tbl.~\ref{#1}}

\newcommand{\Eqn}[1]{Eqn.~\ref{#1}}

\newcommand{\mode}[1]{\underline{\textsc{#1}}\xspace}

\newcommand{\proj}{\textsc{Kaleido}\xspace}

\definecolor{myorange}{RGB}{255, 212, 121}

\newcommand*\circled[2]{\tikz[baseline=(char.base)]{
            \node[shape=circle,fill=black,inner sep=1pt] (char) {\textcolor{#1}{{\footnotesize #2}}};}}

\renewcommand{\hl}[1]{{#1}}

\begin{document}

\title[\proj: Algorithm-Hardware Co-Design for Video Diffusion Transformers]{\proj: Algorithm-Hardware Co-Design for Video Diffusion Transformers by Exploiting Latent Space Correlations}
% \subtitle{\normalsize{MICRO 2026 Submission
%     \textbf{\#55} -- Confidential Draft -- Do NOT Distribute!!}}
%\author{\normalsize{ISCA 2025 Submission
 %   \textbf{\#NaN} -- Confidential Draft -- Do NOT Distribute!!}}

\author{Wenxuan Miao}
\orcid{}
\affiliation{%
  \institution{Shanghai Jiao Tong University}
  \city{Shanghai}
  \country{China}
}
\email{miaowenxuan@sjtu.edu.cn}

\author{Haosong Liu}
\orcid{}
\affiliation{%
  \institution{Shanghai Jiao Tong University}
  \city{Shanghai}
  \country{China}
}
\email{2436824987@sjtu.edu.cn}

\author{Weiming Hu}
\orcid{}
\affiliation{%
  \institution{Shanghai Jiao Tong University}
  \city{Shanghai}
  \country{China}
}
\email{weiminghu@sjtu.edu.cn}

\author{Zihan Liu}
\orcid{0000-0002-0874-0682}
\affiliation{%
  \institution{Shanghai Jiao Tong University, Shanghai Qi Zhi Institute}
  \city{Shanghai}
  \country{China}
}
\email{altair.liu@sjtu.edu.cn}

\author{Aiyue Chen}
\orcid{}
\affiliation{%
  \institution{Huawei Technologies}
  \city{Shanghai}
  \country{China}
}
\email{chenaiyue@huawei.com}

\author{Jianlin Yu}
\orcid{}
\affiliation{%
  \institution{Huawei Technologies}
  \city{Shanghai}
  \country{China}
}
\email{yujianlin1@huawei.com}

\author{Yiwu Yao}
\orcid{}
\affiliation{%
  \institution{Huawei Technologies}
  \city{Shanghai}
  \country{China}
}
\email{yiwuyao@pku.edu.cn}

\author{Yiming Gan}
\orcid{0000-0002-2033-5057}
\affiliation{%
  \institution{ICT, Chinese Academy of Sciences}
  \city{Beijing}
  \country{China}
}
\email{ganyiming@ict.ac.cn}

\author{Jieru Zhao}
\orcid{}
\affiliation{%
  \institution{Shanghai Jiao Tong University}
  \city{Shanghai}
  \country{China}
}
\email{zhao-jieru@sjtu.edu.cn}

\author{Jingwen Leng}
\orcid{0000-0002-5660-5493}
\affiliation{%
  \institution{Shanghai Jiao Tong University, Shanghai Qi Zhi Institute}
  \city{Shanghai}
  \country{China}
}
\email{leng-jw@sjtu.edu.cn}

\author{Minyi Guo}
\orcid{0000-0003-0034-2302}
\affiliation{%
  \institution{Shanghai Jiao Tong University, Shanghai Qi Zhi Institute}
  \city{Shanghai}
  \country{China}
}
\email{guo-my@cs.sjtu.edu.cn}

\author{Yu Feng}
\authornote{Corresponding Author.}
\orcid{0000-0002-2192-5737}
\affiliation{%
  \institution{Shanghai Jiao Tong University, Shanghai Qi Zhi Institute}
  \city{Shanghai}
  \country{China}
}
\email{y-feng@sjtu.edu.cn}

\renewcommand{\shortauthors}{Wenxuan Miao et al.}


%%%%%% -- PAPER CONTENT STARTS-- %%%%%%%%

\begin{abstract}

Video diffusion transformers (vDiTs) generate high-quality video but introduce extremely high compute cost due to the long diffusion timesteps and self-attention computation.
As diffusion timesteps are reduced, the computation cost of self-attention becomes the dominant bottleneck.
Existing acceleration approaches largely inherit sparse attention techniques from large language models, which fail to consider the unique spatio-temporal correlation of video data.

This paper presents \proj, an algorithm–hardware co-design that accelerates all operations in vDiTs by exploiting channel-wise spatio-temporal correlations in latent space.
Based on this insight, we propose a lightweight channel-wise reuse algorithm that skips redundant computations by reusing partial results while preserving higher generative quality than prior methods ($>$17~dB).
To efficiently support this algorithm, we design a systolic-array-like accelerator with reconfigurable processing elements and a lightweight data dispatcher to mitigate irregular sparsity and data access patterns introduced by our reuse algorithm.
Evaluations across three mainstream vDiT models show that \proj achieves up to 5.9$\times$ speedup and 16.0$\times$ energy savings over state-of-the-art accelerators.

\end{abstract}
%\begin{CCSXML}
%<ccs2012>
%  <concept_id>00000000.0000000.0000000</concept_id>
%  <concept_desc>Do Not Use This Code, Generate the Correct Terms for Your Paper</concept_desc>
%  <concept_significance>500</concept_significance>
%  %<concept_id>00000000.00000000.00000000</concept_id>
%  <concept_desc>Do Not Use This Code, Generate the Correct Terms for Your Paper</concept_desc>
%  <concept_significance>300</concept_significance>
%  %<concept_id>00000000.00000000.00000000</concept_id>
%  <concept_desc>Do Not Use This Code, Generate the Correct Terms for Your Paper</concept_desc>
%  <concept_significance>100</concept_significance>
 % <concept_id>00000000.00000000.00000000</concept_id>
%  <concept_desc>Do Not Use This Code, Generate the Correct Terms for Your Paper</concept_desc>
%  <concept_significance>100</concept_significance>
%</ccs2012>
%\end{CCSXML}

%\ccsdesc[500]{Do Not Use This Code~Generate the Correct Terms for Your Paper}
%\ccsdesc[300]{Do Not Use This Code~Generate the Correct Terms for Your Paper}
%\ccsdesc{Do Not Use This Code~Generate the Correct Terms for Your Paper}
%\ccsdesc[100]{Do Not Use This Code~Generate the Correct Terms for Your Paper}


\maketitle

\section{Introduction}
\label{sec:intro}

Large video generative models~\cite{kling, sora, hunyuan, veo2, opensora, lin2024open, hong2022cogvideo, singer2022make, mlgen2, xu2024easyanimate, seawead2025seaweed, zeng2024make, genmo2024mochi, wan2025} have drawn tremendous attention from both academia and industry in the last two years. 
Their ability to generate high-fidelity videos has led to rapid adoption in movie editing~\cite{veo3, xu2025beyond}, advertisement~\cite{coca_ads, predis_ads}, virtual world creation~\cite{worldlab, genie3}, and more.
Beyond their substantial economic potential, these models also offer a unique lens to understand the physical world by learning geometric and causal relationships in natural scenes. 
Thus, they are regarded as a foundation for general-purpose artificial intelligence~\cite{feifeili}.

Today’s mainstream large video generative models are predominantly based on video diffusion transformers (vDiTs).
Similar to other diffusion models, a vDiT progressively reverses a forward noise-adding process, iteratively removing latent noises until a video is clean. 
However, this iterative denoising process is compute-intensive.
For example, generating even a short 5-second 720p video clip with a state-of-the-art vDiT, HunyuanVideo~\cite{kong2024hunyuanvideo}, can take over 30 minutes on a single Nvidia H100 GPU with 80 GB of memory.

This inefficiency stems from two fundamental factors. 
First, the diffusion process consists of many denoising timesteps.
Each requires a full forward inference, resulting in long inference latency.
Second, each inference performs compute-intensive self-attention over all spatio-temporal tokens, making attention the dominant bottleneck as video resolutions or lengths grow.
Recently, the community has made tremendous progress on reducing the number of timesteps via distillation~\cite{hacohen2024LTXVideo, zhang2025turbodiffusion, li2025unleashing, salimans2022progressive}, caching~\cite{zhao2024real, zou2024accelerating, chen2024delta, ma2024deepcache, wimbauer2024cache, liu2024smoothcache}, etc.
As the number of timesteps decreases, self-attention computation becomes the primary bottleneck (\Fig{fig:trend}).

In this paper, we address the compute-intensive self-attention in vDiTs via an algorithm–hardware co-design, \proj.
Unlike prior work~\cite{xi2025sparse, yang2025sparsevideogen2, yoo2024adaptiv, heo2025exion, you2023vitcod}, which largely inherits sparse self-attention optimizations from large language models (LLMs), we introduce a key insight that exploits the \textit{spatio-temporal similarity among latent tokens} to dramatically reduce attention computation by up to 85\%.
The key is to dissect and understand the various patterns of self-attention scores that exist across all vDiT models~\cite{kong2024hunyuanvideo, wan2025, zhang2025turbodiffusion, opensora, yang2024cogvideox, hacohen2024LTXVideo}.
Our experiments show that the cause of these patterns is from the \textit{channel-wise position encoding} used in individual token channels (\Sect{sec:motiv:pattern}).
This encoding mechanism naturally induces spatial and temporal correlations, which accumulate across channels and lead to various patterns in attention scores.

With our key insight, we introduce our \textit{channel-wise reuse algorithm} in \Sect{sec:algo}, which accelerates not only self-attention but also other operations in vDiTs. 
Our algorithm simply serves as a plug-in that identifies the correlations between tokens before each operation, e.g., self-attention, and skips unnecessary computations by reusing previous partial results.
Specifically, our algorithm first evaluates the similarity of adjacent tokens at the token channel level to identify values that can safely share partial computations.
During actual operation, channels with high similarity skip explicit computation and reuse previously computed partial results instead.
Evaluations in \Sect{sec:eval} show that our algorithm saves significant computations while retaining much higher generative quality ($>$17dB) compared to prior algorithm-hardware co-designs~\cite{kong2024cambricon, yoo2024adaptiv, heo2025exion}.

While our algorithm significantly reduces overall computation, it introduces two key challenges. 
First, its reuse patterns are inherently incompatible with the dataflow of existing accelerators, e.g., systolic arrays. 
Second, reuse further induces irregular sparsity in the input data. 
To address these challenges, we co-design a reconfigurable processing element (PE) for systolic arrays, as described in \Sect{sec:arch:dataflow}. 
Our PE can be reconfigured into multiple execution modes and supports various dataflows, so that PEs can reuse the previously computed results without repeatedly loading them from on-chip memory. 
In addition, it incorporates mixed-precision computation to support computations with partially reused sparse inputs.

To further mitigate the irregular data access patterns introduced by our reuse algorithm, we propose a data dispatcher in \Sect{sec:arch:dispatcher}. 
The dispatcher applies a lightweight online clustering to group tokens with similar reuse patterns.
Meanwhile, a runtime lookahead buffer is designed to merge compatible channels before feeding them into the PE array.
Together, we show that our data dispatcher further improves PE utilization and overall accelerator efficiency (\Sect{sec:arch:dispatcher}).

We implement our algorithm-hardware co-design, \proj. 
The hardware builds on top of a classic systolic DNN accelerator~\cite{jouppi2017datacenter} implemented in 16~nm technology. 
The \proj hardware augments the baseline accelerator with 7.9\% area overhead. 
We evaluate \proj on four widely-adopted vDiTs: HunyuanVideo~\cite{kong2024hunyuanvideo}, Wan~\cite{wan2025}, CogVideoX~\mbox{\cite{yang2024cogvideox}} and TurboDiffusion~\cite{zhang2025turbodiffusion}.
In addition to the GPU baseline, we also include four recent diffusion accelerators~\cite{kong2024cambricon, yoo2024adaptiv, heo2025exion, kim2025ditto}.
\proj achieves up to 5.9$\times$ speedup and 16.0$\times$ energy savings. 
The contributions of this paper are as follows:
\begin{itemize}
    \item We are the first to systematically characterize the large vDiT models and explain the spatio-temporal correlations in the latent space of large vDiTs.
	\item We propose a lightweight channel-wise reuse algorithm that significantly reduces computation while achieving higher generative quality than prior methods.
    \item We introduce a reconfigurable accelerator architecture to support our reuse algorithm and a tailored data dispatcher to mitigate irregular sparsity and improve PE utilization.
    \item Our architecture achieves up to 5.9$\times$ speedup and 16.0$\times$ energy savings over existing accelerator designs.
\end{itemize}

\section{Background}
\label{sec:bg}

\begin{figure}
    \centering
    \includegraphics[width=\columnwidth]{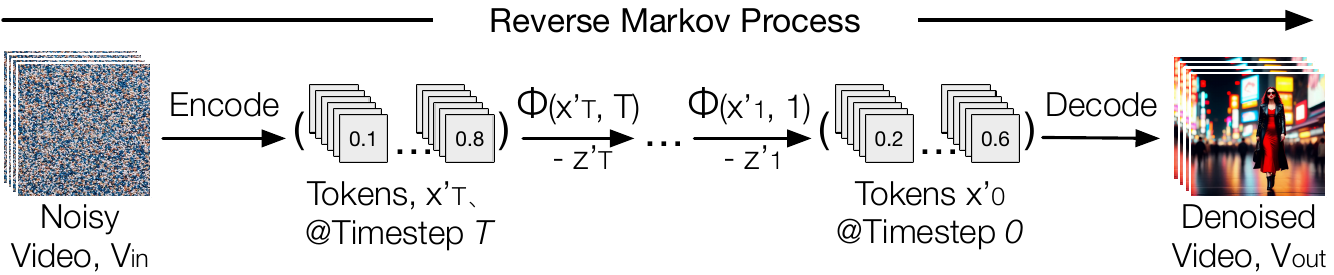}
    \caption{An example of a diffusion process. The random-noised video input, $V_{in}$, is first encoded into the latent space and converted into a sequence of tokens, $x'_{T}$. These tokens are then processed by the same diffusion model multiple times to predict the noises, $z'_t$, at each timestep, $t$. After completing $T$ timesteps, the final denoised tokens, $x'_0$, are decoded back into a final output video, $V_{out}$, in RGB color space.}
    \label{fig:diffusion_process}
\end{figure}

\para{Diffusion Models.}
The most widely adopted video generation paradigm is the diffusion model, which learns to generate data, i.e., images or videos, from random Gaussian noise through a reverse Markov process~\cite{croitoru2023diffusion, yang2023diffusion, ho2020denoising}, as shown in \Fig{fig:diffusion_process}. 

In this process, the initial input is a random Gaussian noise, $V_{in}$, which is encoded into tokens, $x'_T$, in latent space.
Here, $T$ is the total number of denoising timesteps.
Given noisy tokens $x'_T$, the diffusion model restores the original data by gradually predicting and removing noise $z'_t$ from $x'_t$ at each timestep $t$.
After a fixed number of denoising timesteps, the final result of the diffusion model, $x'_0$, would be close enough to the original data $x_0$ in latent space.
The mathematical expression of each denoising timestep is,
\begin{equation}
    x'_{t-1} = \alpha_t (x'_{t} - \beta_t z'_t) +  \sigma_t n'_t,\ \ \text{and} \ z'_t = \Phi(x'_t, t),
\end{equation}
where $\Phi$ is the prediction function, i.e., the diffusion model, that predicts the noise $z'_t$.
Both $\alpha_t$ and $\beta_t$ are hyper-parameters.
$\sigma_t n'_t$ is a renoising term to add randomness to the denoising process.
Lastly, we decode $x'_0$ from the latent space to obtain the output video, $V_{out}$.

Current mainstream diffusion models in video generation often require 30 to 100 denoising steps to generate a short video~\cite{wan2025, genmo2024mochi, opensora, lin2024open, hong2022cogvideo, yang2024cogvideox, kong2024hunyuanvideo, meituan2025longcatvideo}.
Recent studies also reduce to shorter timesteps, 4-10, via distillation~\cite{liu2025infinityStar, zhang2025turbodiffusion}, finetuning~\cite{hacohen2024LTXVideo, liu2025astraea, liu2024timestep}, etc.

% \begin{figure}
%     \centering
%     \includegraphics[width=\columnwidth]{exe_breakdown_rtxpro6000}
%     \caption{The normalized execution time of different vDiTs on RTX pro 6000. Self-attention dominates the overall execution. (\fixme{fake})}
%     \label{fig:exe_breakdown}
% \end{figure}

% \begin{figure}
%     \centering
%     \includegraphics[width=\columnwidth]{exe_breakdown_h100}
%     \caption{The normalized execution time of different vDiTs on H100. Self-attention dominates the overall execution. (\fixme{fake})}
%     \label{fig:exe_breakdown}
% \end{figure}

\begin{figure}[t]
    \centering
    \subfloat[Nvidia H100 with HBM~\cite{h100}.]{
        \includegraphics[width=0.46\columnwidth]{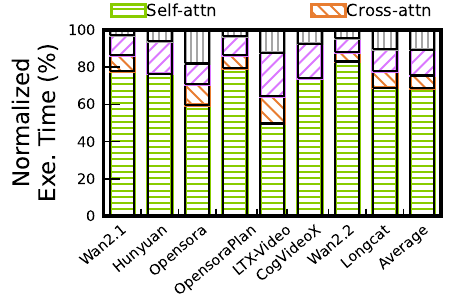}
        \label{fig:breakdown_h100}
    }
    \vspace{1pt}
    \subfloat[Nvidia RTX PRO 6000~\cite{rtx6000pro}.]{
        \includegraphics[width=0.46\columnwidth]{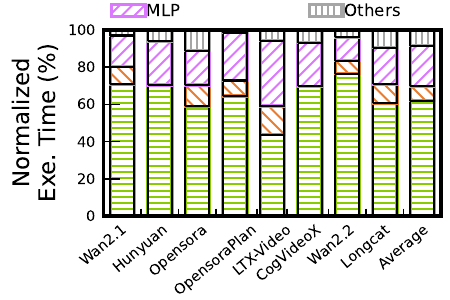}
        \label{fig:breakdown_rtx6000}
    }
    \caption{The execution breakdown of eight mainstream  vDiTs~\cite{lin2024open, kong2024hunyuanvideo, hong2022cogvideo, wan2025, meituan2025longcatvideo, hacohen2024LTXVideo, opensora} on two recent Nvidia GPUs~\cite{h100, rtx6000pro}. Self-attention dominates the overall execution.}
    \label{fig:breakdown}
\end{figure}

\begin{figure}[t]
    \centering
    \subfloat[Nvidia H100 with HBM~\cite{h100}.]{
        \includegraphics[width=0.46\columnwidth]{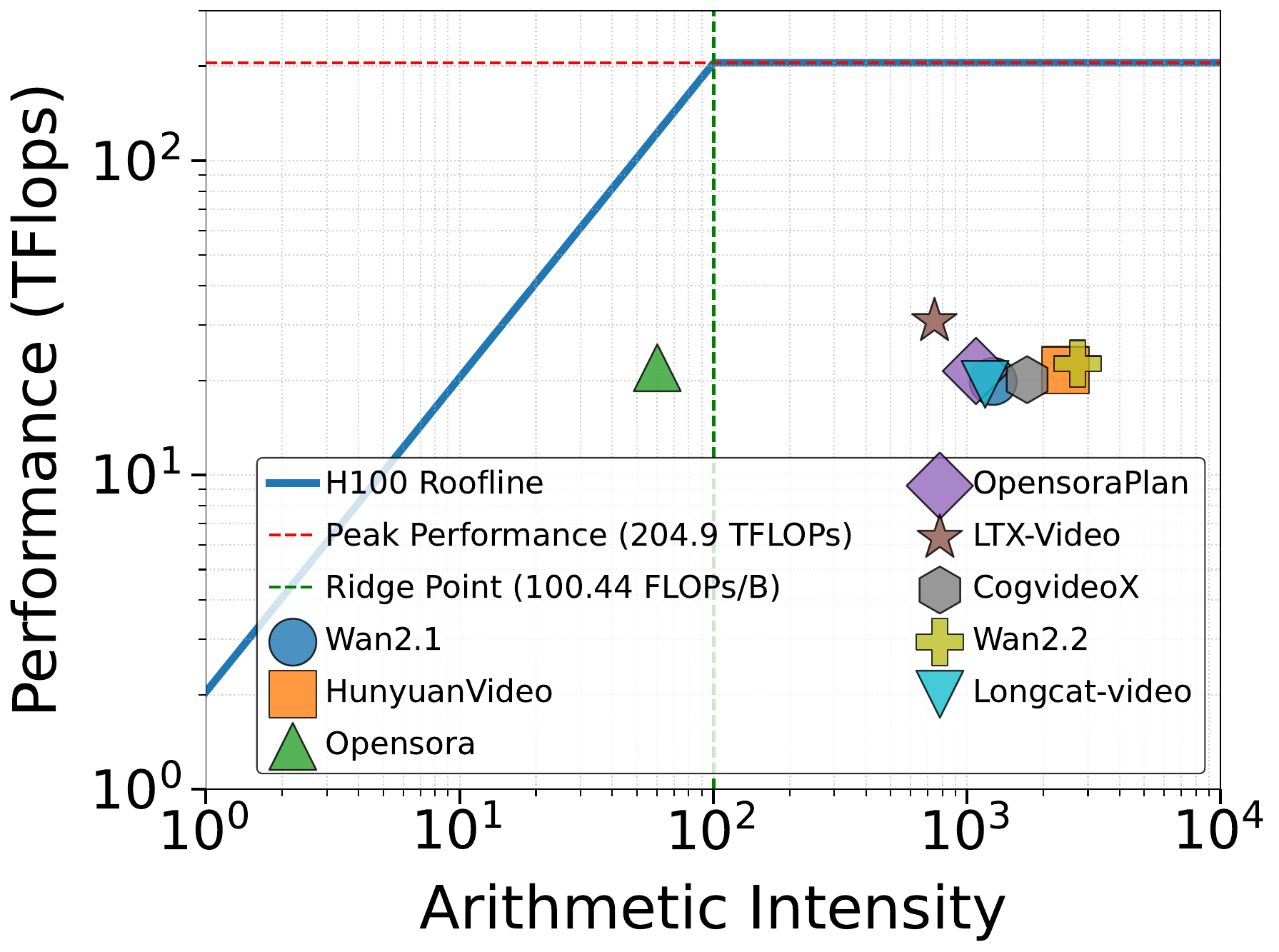}
        \label{fig:h100}
    }
    \vspace{1pt}
    \subfloat[Nvidia RTX PRO 6000~\cite{rtx6000pro}.]{
        \includegraphics[width=0.46\columnwidth]{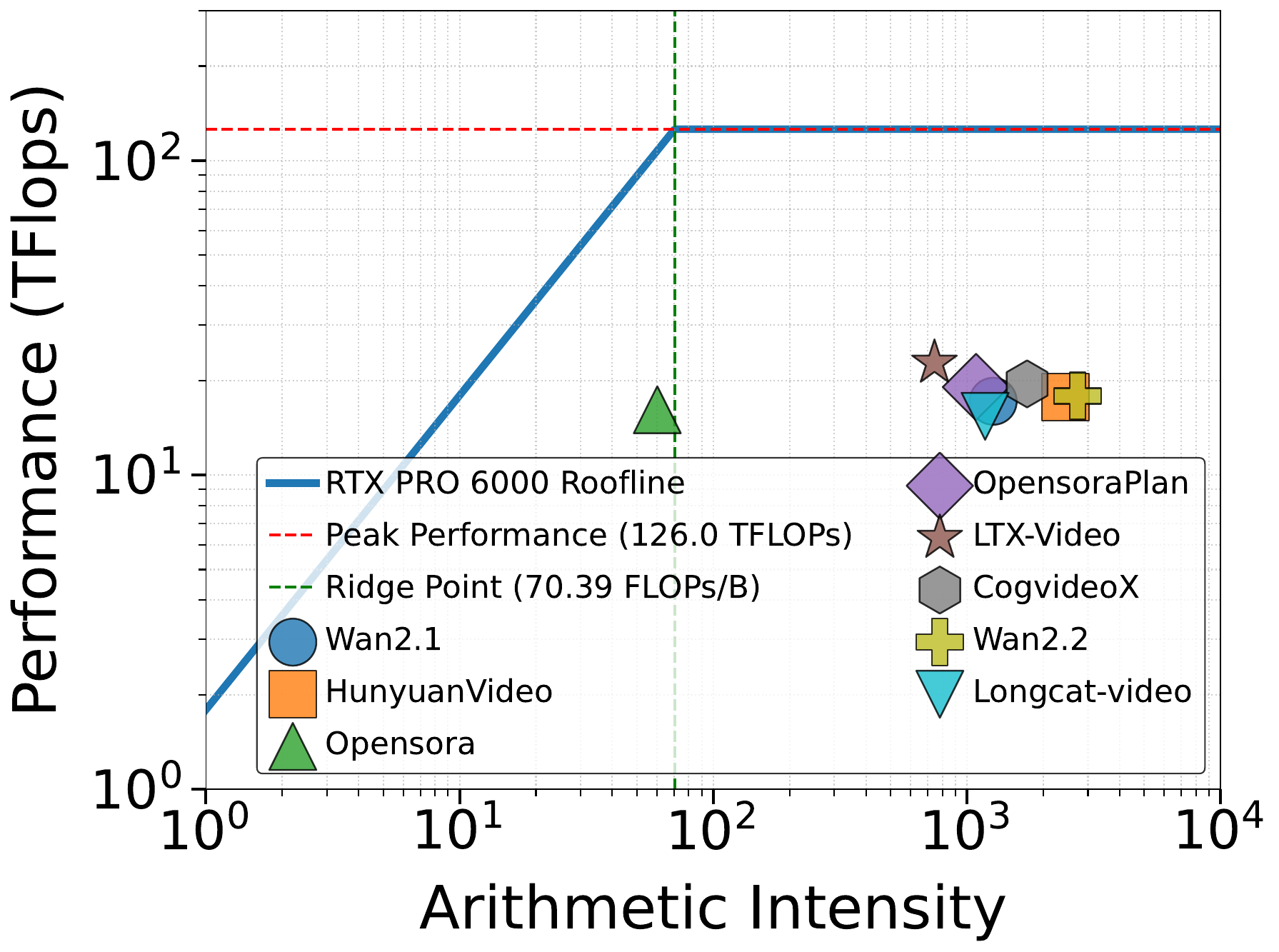}
        \label{fig:rtx6000}
    }
    \caption{The roofline analyses of eight vDiT models~\cite{lin2024open, kong2024hunyuanvideo, hong2022cogvideo, wan2025, meituan2025longcatvideo, hacohen2024LTXVideo, opensora} on two recent Nvidia GPUs~\cite{h100, rtx6000pro}. All vDiT models are primarily compute-bound.}
    \label{fig:roofline}
\end{figure}

\para{Video Diffusion Transformer.}
A typical vDiT architecture first encodes input and prompt tokens into latent space with $d$ channels. 
Then, for each timestep, those encoded tokens go through a stack of computational blocks.
Each block is a combination of a self-attention, a cross-attention, and a MLP layer.
\Fig{fig:breakdown} profiles the execution time of eight popular vDiT models~\cite{lin2024open, kong2024hunyuanvideo, hong2022cogvideo, wan2025, meituan2025longcatvideo, hacohen2024LTXVideo, opensora}.
Overall, self-attention accounts for 68\% of total execution time.

\para{Self-Attention.}
The attention mechanism is the main reason that self-attention is the major computational bottleneck in vDiTs.
Given input tokens $X_\text{in} \in \mathbb{R}^{N \times d}$, self-attention computes,
\begin{equation}
    \text{Attention}(Q, K, V) = \text{Softmax}\left(\frac{QK^T}{\sqrt{d_k}}\right)V,
    \label{eqn:attn}
\end{equation}
where $Q, K, V$ are query, key, and value, respectively. 
$N$ is the token sequence and $d$ is the channel dimension.
These three token sequences have the same shape as the input tokens and are obtained by performing linear projections on $X_\text{in}$.
$P = QK^T \in \mathbb{R}^{N \times N}$ is the attention map.
Overall, the computational complexity of self-attention is proportional to $N^2\times d$.
Thus, it is compute-intensive.
\Fig{fig:roofline} shows the roofline analyses of vDiTs on Nvidia H100~\cite{h100} and Nvidia RTX PRO 6000~\cite{rtx6000pro}.
Under two different memory technologies, HBM2e and GDDR7, vDiT models are still compute-bound.
Here, Opensora~\cite{opensora} is not compute-bound because it uses temporal-spatial separate attention, which is no longer used in recent vDiTs.

\para{Position Encoding.}
Next, we explain one of the key concepts in vDiT.
The reason that vDiT models can capture spatial-temporal information is that all vDiTs apply rotary positional embedding (RoPE)~\cite{lin2024open, hunyuan, hong2022cogvideo, wan2025} to encode relative token positions,
\begin{equation}
    \text{RoPE}([x, y]) = (\cos\theta \cdot x - \sin\theta \cdot y,\ \sin\theta \cdot x + \cos\theta \cdot y),
\end{equation}
where $\theta_{p,i} = \varphi / 10000^{2i/d}$. Importantly, the $d$ channels are often partitioned into three segments: initial channels encode temporal information, while the remaining channels capture spatial information along $x$ and $y$ axes. 
This structured encoding causes different channel groups to exhibit unique spatio-temporal patterns.

\section{Motivation}
\label{sec:motiv}

\subsection{Trends and Challenges of vDiT Models}
\label{sec:motiv:ch}

\begin{figure}
    \centering
    \includegraphics[width=\columnwidth]{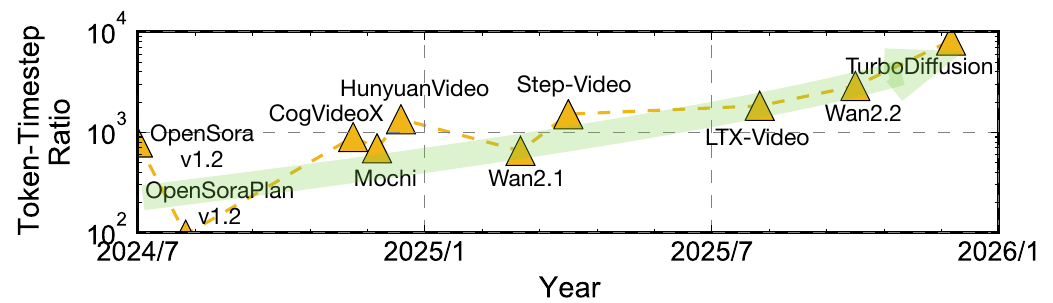}
    \caption{The ratio between token sequence length and the number of diffusion timesteps in vDiTs~\cite{opensora, lin2024open, yang2024cogvideox, genmo2024mochi, kong2024hunyuanvideo, wan2025, ma2025stepvideot2v, hacohen2024LTXVideo, zhang2025turbodiffusion} has steadily increased over the years. 
    It means that attention computation grows to be the only bottleneck.}
    \label{fig:trend}
\end{figure}

As mentioned in \Sect{sec:bg}, the inference cost of a vDiT is primarily dominated by two factors: the number of denoising timesteps and the computation of self-attention.
While both factors have historically posed major bottlenecks, recent studies~\cite{chen2024delta, liu2024timestep, zou2024accelerating, zhao2024real, liu2025astraea, chen2025dove, zhang2025turbodiffusion, hacohen2024LTXVideo} exploit various techniques to leverage the similarity between timesteps and reduce the number of timesteps by approximating intermediate results. 
As these techniques become widely adopted, the diffusion timestep is no longer a dominant bottleneck.

In contrast, \textit{self-attention remains a key bottleneck}, especially as recent vDiTs target longer video durations and higher resolutions~\cite{liu2025infinityStar, meituan2025longcatvideo, chen2025sana, zhang2025turbodiffusion, hunyuanworld2025tencent, lingbot-world}.
Both trends translate into longer token sequences, causing a quadratic increase in self-attention cost.
As shown in \Fig{fig:trend}, we show that the ratio between the token sequence and the number of timesteps across mainstream vDiTs increases over time.
This means that accelerating self-attention will be the key to improving the performance of future vDiTs.

While a substantial body of prior studies~\cite{ye2025flashinfer, dao2022flashattention, dao2023flashattention, wu2023highlight, zhang2020sparch, wang2021spatten, lu2021sanger, ham2021elsa} have studied sparse computation to accelerate self-attention in LLMs, these approaches primarily exploit attention sparsity to eliminate insignificant computation.
Several recent works on image and video diffusion models~\cite{yoo2024adaptiv, heo2025exion, xi2025sparse, you2023vitcod} have adopted similar ideas. 
However, as we show in \Sect{sec:eval:quality}, directly transferring LLM-based sparse attention techniques to vDiTs is both ineffective and fundamentally mismatched to the characteristics of vDiT models.

Sparse attention is intuitive for LLMs because text sequences naturally exhibit discrete, hierarchical semantics.
Thus, many long-range irrelevant dependencies can be safely pruned without any impact on generation quality.
Unlike text tokens, video tokens are dense and carry continuous visual semantics.
Every token contributes to spatial coherence and temporal consistency, especially in early denoising steps where global information must be retained to achieve consistent results~\cite{zhao2024real, xi2025sparse, chen2024delta}.
Our experiment in \Sect{sec:eval:quality} shows that naively removing attention correlations leads to motion discontinuities and flickering artifacts. 
Thus, rather than relying on sparsity patterns in LLMs, effective acceleration of self-attention in vDiTs should exploit their unique spatio-temporal correlations.

\subsection{Spatio-Temporal Correlations in vDiTs}
\label{sec:motiv:pattern}

In this subsection, we first present the attention patterns that commonly exist in vDiTs, and then explain the underlying causes that induce these spatio-temporal correlations in their latent space.

\begin{figure}
    \centering
    \includegraphics[width=\columnwidth]{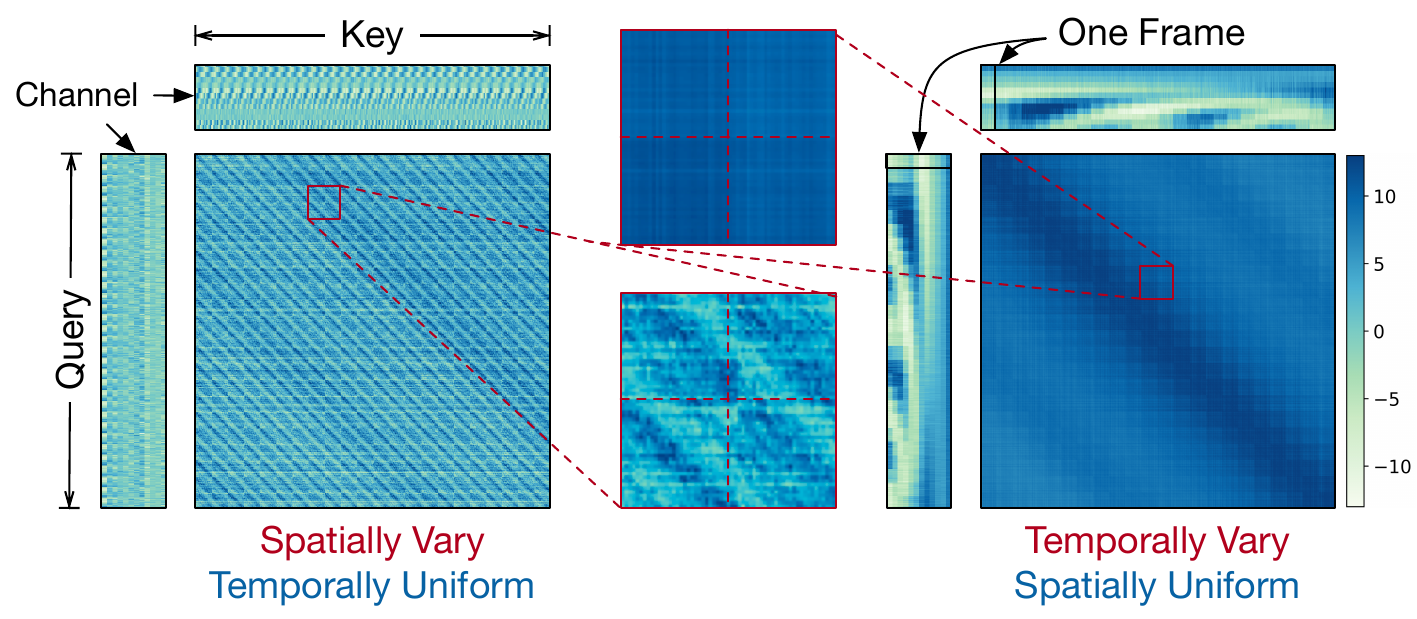}
    \caption{Example of various attention patterns, which can be classified into \textit{spatial patterns} and \textit{temporal patterns}. Spatial patterns capture the spatial correlations within a single frame by repetitive ``tiles'', whereas temporal patterns capture the temporal correlations across frames via strong diagonal correlations. 
    Both patterns arise from the RoPE encoding and the dominant channels in query $Q$ and key $K$.
    }
    \label{fig:attn_patterns}
\end{figure}

\para{Spatio-Temporal Patterns.}
In \Fig{fig:attn_patterns}, we show two representative self-attention patterns from HunyuanVideo~\cite{kong2024hunyuanvideo}, other vDiT models have similar patterns.
Also, the patterns in \Fig{fig:attn_patterns} are consistent with patterns reported in prior studies~\cite{xia2025training, yuan2024ditfastattn, ding2025efficient, xi2025sparse}.
Overall, the patterns in self-attention maps of vDiTs can be categorized into two types: \textit{spatial patterns} and \textit{temporal patterns}.

An example of a spatial pattern is shown on the left side of \Fig{fig:attn_patterns}.
Spatial patterns consist of small, seemingly ``repetitive'' tiles spanning across the entire attention map, $P$. 
Each tile captures the spatial correlations among tokens within a single frame, while the patterns across different tiles remain largely unchanged.
Thus, spatial patterns mainly capture spatial relationships within a frame.

In contrast, temporal patterns, shown on the right side of \Fig{fig:attn_patterns}, emphasize correlations across frames.
In these patterns, the variations of attention scores within a single frame are less significant, whereas inter-frame correlations are more important.
In particular, temporally adjacent frames tend to exhibit strong mutual attention, resulting in a diagonal correlation in the attention map.

\para{Underneath Mechanism.}
While prior studies~\cite{xi2025sparse, yang2025sparsevideogen2} also observed these patterns in vDiTs, they do not further investigate the underlying cause of these patterns.
Existing approaches simply leverage these patterns to accelerate vDiTs via sparse computations of attention maps using the techniques from LLMs.
However, without a fundamental understanding of the underlying mechanisms, these heuristics often lead to low generative quality.
% \Sect{sec:eval:quality} compares the quality of our method against prior works.

\begin{figure}
    \centering
    \includegraphics[width=\columnwidth]{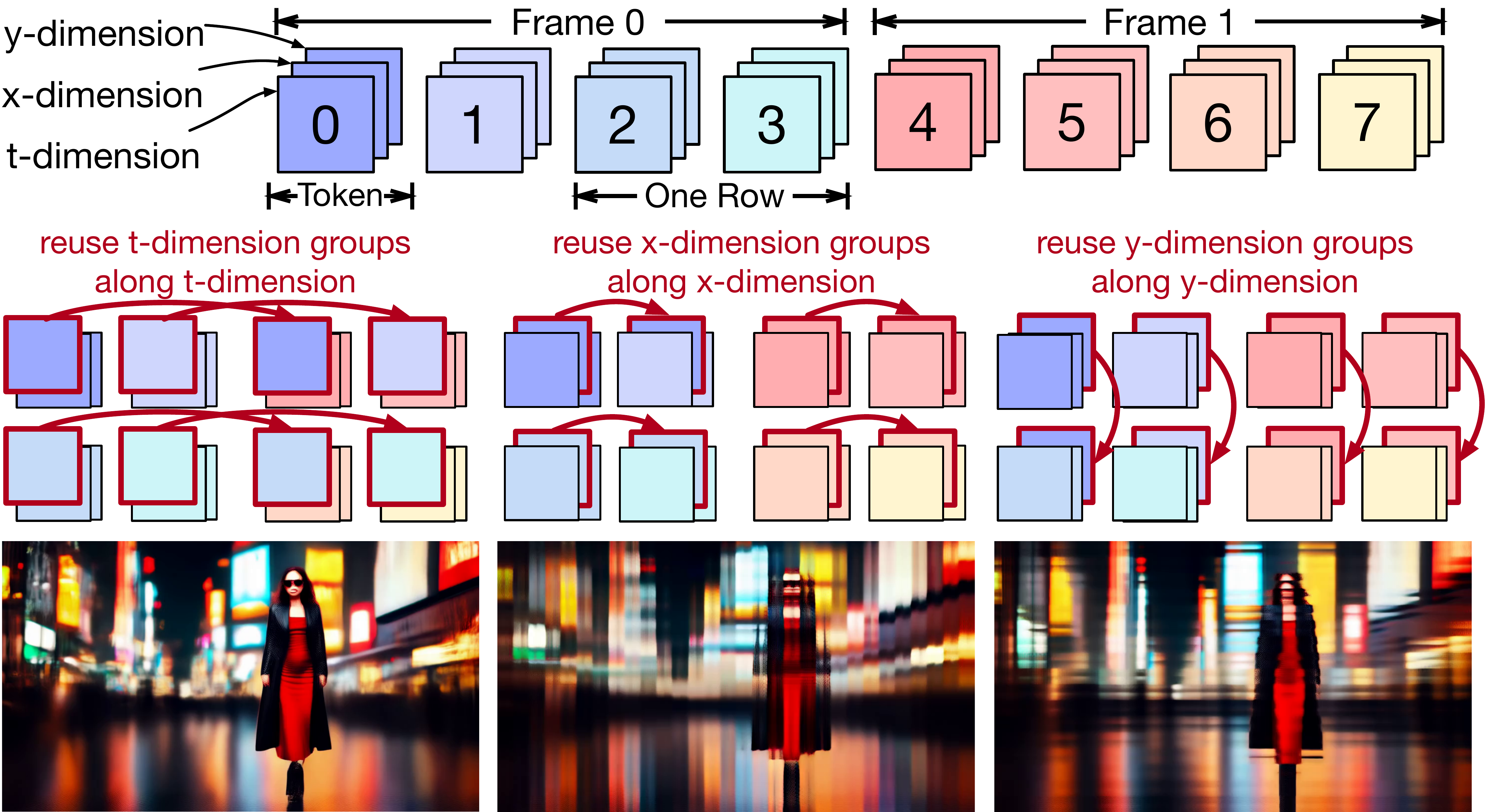}
    \caption{An example of how different channel groups govern the final generation quality. The upper part illustrates how we reuse different channel groups shown by red arrows. Three channels represent t-, x-, and y-dimension channel groups. The lower part shows results after reusing.}
    \label{fig:reuse_example}
\end{figure}

In this work, we take the first step to explain the causes of various attention patterns and propose a principled acceleration technique tailored for vDiTs.
Our analysis shows that the dominant driver of those patterns is the \textit{channel-wise encoding} of tokens, i.e., RoPE.
As introduced in \Sect{sec:bg}, RoPE is the primary positional encoding technique in vDiTs, where different channel groups encode positional information at different frequencies.
Specifically, vDiT models partition their channel dimensions into three distinct groups: the temporal (t) dimension, the horizontal (x) dimension, and the vertical (y) dimension, as shown in \Fig{fig:reuse_example}.
Each channel group captures information along a specific direction in latent video data.
\Tbl{tab:model_config} lists the channel partitioning of vDiT models.

\begin{table} 
\caption{The channel partitioning of different vDiT models~\cite{kong2024hunyuanvideo, wan2025, zhang2025turbodiffusion, yang2024cogvideox, meituan2025longcatvideo, lin2024open}. vDiT models apply RoPE encoding and partition the token channels into t-, x-, and y-dimension.}
\centering
\resizebox{\linewidth}{!}{
\renewcommand*{\arraystretch}{1}
\renewcommand*{\tabcolsep}{3pt}
\begin{tabular}{ c|c|c|c|c|c|c } 
\toprule[0.15em]
 & HunyuanVideo & Wan2.1 & TurboDiffusion & CogVideoX & LongCat & OpenSoraPlan \\ 
\midrule[0.05em]
T-dimension & 16 & 44 & 44 & 32 & 44 & 32 \\ 
X-dimension & 56 & 42 & 42 & 48 & 42 & 32 \\ 
Y-dimension & 56 & 42 & 42 & 48 & 42 & 32 \\ 
\bottomrule[0.15em]
\end{tabular}
}
\label{tab:model_config}
\end{table}

Meanwhile, we find that the dominant channels with higher values have a higher chance to contribute more to the attention scores in \Fig{fig:attn_patterns}.
Here, the bounding boxes on $Q$ and $K$ highlight individual frames.
Because RoPE causes the dominant channels to have different frequencies, leading to various patterns in $Q$ and $K$.
Different combinations of channels in $Q$ and $K$ form various patterns in attention maps.
As shown in \Fig{fig:attn_patterns}, when spatial channels dominate, attention maps exhibit ``spatial-like'' patterns.
Otherwise, attention maps exhibit ``temporal-like'' patterns.

\para{Theory Verification.}
To verify our theory, we design an experiment that purposely manipulates different channel groups and examines their impact on the final video quality.
As shown in \Fig{fig:reuse_example}, for each channel group, we group pairs of adjacent tokens along the corresponding dimension and force the second token in each pair to reuse the value of the first token across all denoising steps.
For example, for the temporal channel group, we allow the t-dimension channels of every second frame to reuse those of the first frame at the same coordinate, while leaving the x- and y-channel groups unchanged, and vice versa.
\Fig{fig:reuse_example} illustrates how we reuse different channel groups with a toy example of eight input tokens (two frames with $2\times2$ tokens per frame).

\begin{figure}[t]
    \centering
    \subfloat[T-channel group.]{
        \includegraphics[width=0.32\columnwidth]{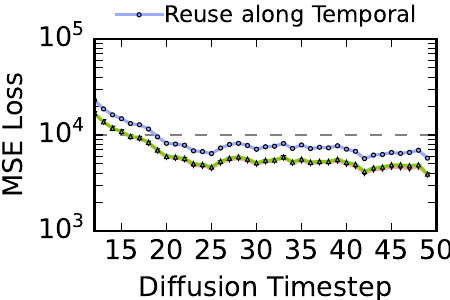}
        \label{fig:t_axis}
    }
    \subfloat[X-channel group.]{
        \includegraphics[width=0.32\columnwidth]{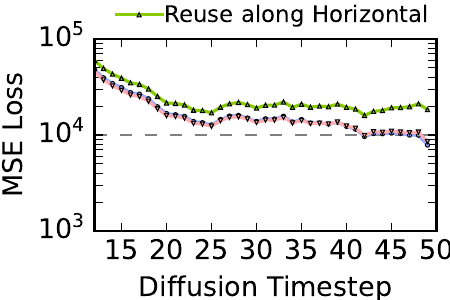}
        \label{fig:x_axis}
    }
    \subfloat[Y-channel group.]{
        \includegraphics[width=0.32\columnwidth]{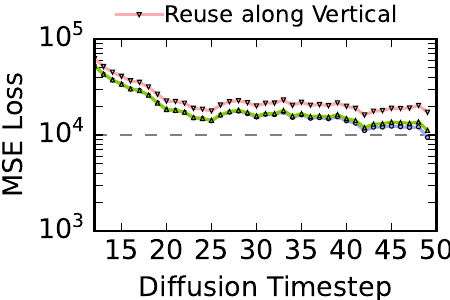}
        \label{fig:y_axis}
    }
    \caption{The MSE loss introduced by reusing different directions on different channel groups. For a given channel group, reuse along the other two dimensions leads to lower MSE loss than reuse along its own direction.}
    \label{fig:axis_loss}
\end{figure}

\begin{figure*}
    \centering
    \includegraphics[width=\linewidth]{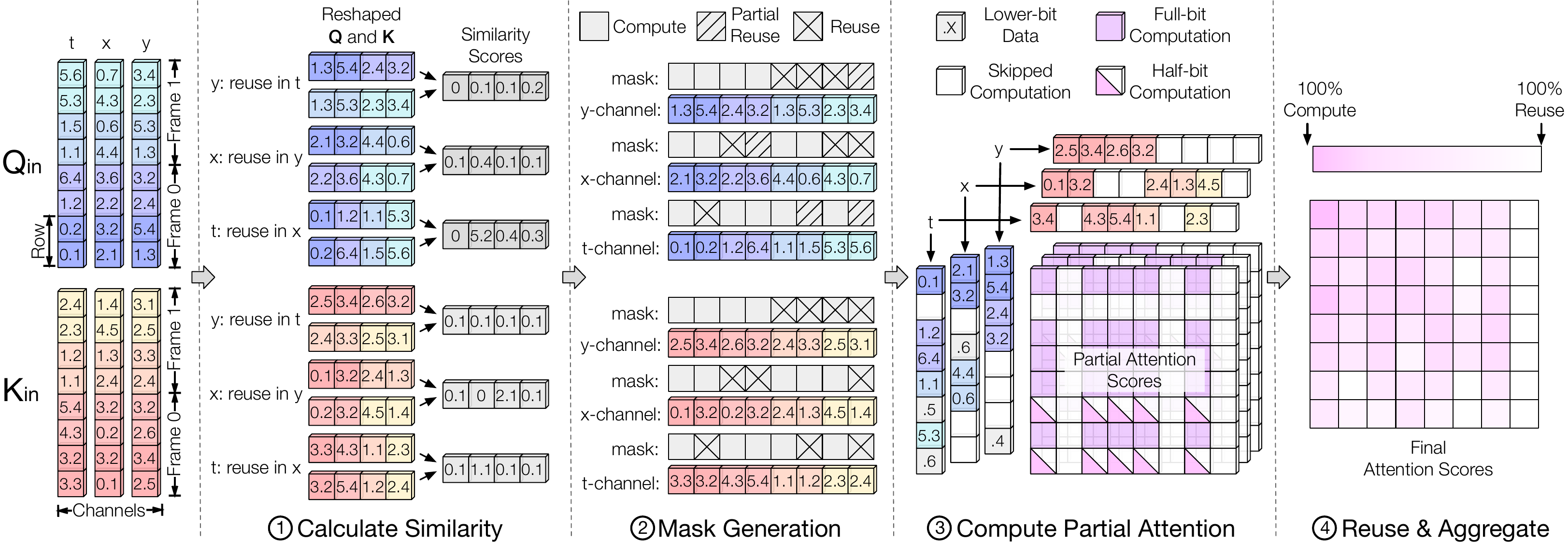}
    \caption{An overview of our channel-wise reuse algorithm, which consists of four steps. We first compute the similarity between adjacent tokens along a selected axis. 
    Based on this similarity, a subset of tokens could reuse previously computed partial attention scores, thus reducing the overall attention computation. $\theta_\text{th1}$ and $\theta_\text{th2}$ are set to be 0.1 and 1 for illustration purposes.}
    \label{fig:idea}
\end{figure*}

The bottom part of \Fig{fig:reuse_example} shows some results after applying our reuse method.
As shown, reusing t-dimension channels introduces temporal distortions, whereas reusing x- or y-dimension channels produces stripe-like artifacts aligned with the corresponding direction.
This shows that different channel groups govern different dimensions of information.
Overall, all used frames are much blurrier than the baseline because the self-attention mechanism integrates global contributions across all tokens.

\Fig{fig:axis_loss} further quantifies the mean square error (MSE) loss between the baseline results and the results generated with different channel reuses.
The x-axis shows the diffusion timesteps at which we apply the reusing strategy.
For every channel group, we evaluate reuse along three different dimensions: t, x, and y.
The result shows that, for a given channel group, reuse is less harmful when applied along the other two dimensions, rather than along the dimension that the channel group encodes.
For instance, for the t-dimension channel group, applying reuse along the x- or y-dimension results in lower MSE than reusing along the t-dimension.
This result further confirms that each channel group predominantly captures information specific to its corresponding dimension.
\section{Channel-Wise Reuse Algorithm}
\label{sec:algo}

With the major finding of the spatio-temporal correlations in \Sect{sec:motiv}, we introduce our \textit{channel-wise reuse algorithm} to accelerate all computation blocks in vDiTs.
Here, we first provide an overall idea of our algorithm in \Sect{sec:algo:idea} and then explain the rationale and design decisions behind our algorithm in \Sect{sec:algo:design}.

\subsection{Idea}
\label{sec:algo:idea}

\Sect{sec:motiv} shows that the spatio-temporal patterns in attention maps are governed by dominant token channels, and each channel has a unique frequency governed by RoPE encoding (see \Fig{fig:attn_patterns}).
Here, we show how to leverage this unique insight to accelerate the operations in vDiTs via reusing intermediate results.
Our algorithm assumes an 8-bit fixed-point numerical representation, which is widely adopted in diffusion accelerators~\cite{kong2024cambricon, kim2025ditto}.
Here, we use self-attention to demonstrate our idea first.
Other compute blocks, i.e., across-attention and MLP, can be accelerated in a similar fashion.

\para{Self-Attention.} 
Our overall process to accelerate self-attention is shown in \Fig{fig:idea}, which consists of four steps.

\circled{white}{1}
Given the mathematical expression of self-attention in \Eqn{eqn:attn}, we first compute the similarity of adjacent tokens in the query $Q$ and key $K$.
Later, these results will guide our reuse algorithm.
As shown in \Fig{fig:axis_loss}, for a given channel group, it is better to reuse the channel values along the other two directions.
For example, channels belonging to the t-channel group have lower MSE loss when reusing tokens along x- or y-directions.
To better co-design with our architecture, we further restrict each channel group to reuse along one fixed direction.
E.g., in \Fig{fig:axis_loss}, t-, x-, and y-channels are restricted to reuse x-, y-, and t-directions, respectively.
At runtime, the token similarity is computed along the predefined direction.
The similarity $\Delta$ is measured using the absolute difference,
\begin{equation}
    \Delta(x_{a}, x_{b}) = |x_{a}-x_{b}|,
\end{equation}
where $x_{a}$ and $x_{b}$ are single-channel values of two adjacent tokens along a specific direction.
For instance, along the t-direction, we compute the similarity $\Delta$ between every two adjacent frames.

\circled{white}{2}
We then check whether the token similarity $\Delta$ is below our predefined thresholds, $\theta_\text{th1}$ and $\theta_\text{th2}$.
Here, $\theta_\text{th1} < \theta_\text{th2}$.
If $\Delta$ is below $\theta_\text{th1}$, then the second token $x_{b}$ is marked as ``\textit{reuse}'', i.e., it can completely reuse the partial attention score of the first token $x_{a}$.
If $\Delta$ is below $\theta_\text{th2}$, the second token $x_{b}$ is then marked as ``\textit{partial reuse}'', i.e., it can partially reuse the attention score of $x_{a}$.

\circled{white}{3}
Once we identify reusable tokens in each channel group, we start to compute the partial attention scores.
Here, $Q$ and $K$ perform matrix multiplication channel by channel.
If any tokens in $Q$ and $K$ are marked as ``reuse'', they will not perform computation and instead reuse previously computed results, as the white blocks shown in \Fig{fig:idea}.
If any tokens in $Q$ and $K$ are marked as ``partial reuse'', they would perform half multiplication.
Consider a fixed-point representation, we could reuse the higher $h$ bits multiplication from the previous token and only compute the subsequent 4 bits, as the half-colored blocks shown in \Fig{fig:idea}.

\circled{white}{4}
Finally, all partial attention scores from all channels are aggregated to calculate the final attention scores $P$.
Once the final attention scores are obtained, operations, e.g., Softmax, are performed the same as in the canonical self-attention.
Lastly, $P\times V$ is performed similarly to $Q\times K^T$. 
However, we only perform a reuse check on $V$, not $P$.
Because only $V$ has spatio-temporal correlations among its tokens.
The reuse process of $V$ is similar to $Q$ and $K$.

\para{Other Operations.}
In addition to self-attention, other operations in vDiTs can also benefit from a similar reuse strategy. 
However, unlike self-attention, for other operations such as MLPs, reuse can only be applied to activations rather than weights.
This is because only activations inherit the spatio-temporal correlations, whereas weights do not.
Thus, for these operations, we modify step \circled{white}{1} of our reuse algorithm; other steps remain largely unchanged.

In step \circled{white}{1}, we first compute the similarity of activations on a per–channel-group basis.
Similar to self-attention, all channels in each channel group are restricted to reuse along one direction.
However, since only one input, e.g., activation, has spatial-temporal correlations, we propose a window-based comparison, rather than comparing two adjacent frames, to increase the reuse ratio.
The similarity $\Delta$ is measured as,
\begin{equation}
\begin{aligned}
    \Delta(a) = \sqrt{\sum_{i=0}^{w-1}(a_i-\bar{a})^2/w},\quad \bar{a}=\sum_{i=0}^{w-1}a_i/w,
\end{aligned}
\end{equation}
where $a$ is a window of activation elements along one channel. 
$w$ is the window size.
$\Delta(a)$ determines whether all elements within this window can be reused or partially reused.
In \Sect{sec:arch:dataflow}, we will show that this window-based design can also map naturally onto our co-designed PE array.

\subsection{Design Decisions}
\label{sec:algo:design}

Next, we explain several key design rationales in our algorithm.

\begin{figure}[t]
\centering
\begin{minipage}[t]{0.48\columnwidth}
  \centering
  \includegraphics[width=\columnwidth]{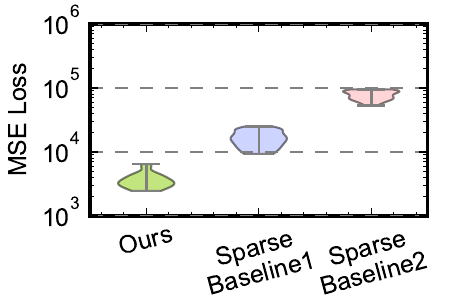}
  \caption{MSE comparison between our reusing method and two skipping techniques. We pick 20 random prompts.}
  \label{fig:reuse_cmp}
\end{minipage}
\hspace{2pt}
\begin{minipage}[t]{0.48\columnwidth}
  \centering
  \includegraphics[width=\columnwidth]{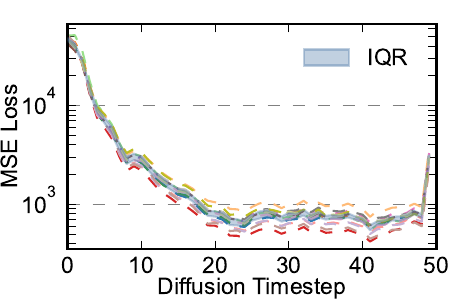}
  \caption{The sensitivity of our reuse technique accuracy to different prompts. Here, we show 20 random prompts.}
  \label{fig:threshold_prompt}
\end{minipage}
\end{figure}

\para{Why Reuse?}
We first explain why our reuse technique is better than conventional sparse attention techniques~\cite{xi2025sparse, yang2025sparsevideogen2, heo2025exion}.
Recall, sparse attention exploits the insignificance of $Q K^T$ values by directly skipping their attention computations.
In \Fig{fig:reuse_cmp}, we compare the MSE loss of our method against two sparse attention baselines under the same token-saving ratio, $\theta^* = 85 \%$, on HunyuanVideo~\cite{hunyuan}.
The first baseline skips 85\% of tokens with the lowest values during the attention computation.
The second baseline uses the same selection criteria as our reuse method to identify tokens, but instead of reusing previously computed results, it skips their attention computations.
The results show the output MSE between the original model and different compute-saving techniques.
Results show that our technique achieves an order of magnitude lower MSE loss compared to these two baselines.
This shows that our technique is more effective than sparse computation methods.
% Because sparse attention techniques are essentially similar to ``reusing'' insignificant tokens that are close to ``0'', while our reuse strategy reuses across the entire value range.

\para{Why Channel?}
Next, we explain why our reuse strategy operates at the channel level rather than the token level, as proposed by prior works~\cite{bolya2023token, yoo2024adaptiv}. 
Prior works target high-level tasks, like classification.
For those tasks, aggregating token information is acceptable.
However, video generation requires preserving dense and smooth content at the pixel level. 
Merging or reusing at the token level often leads to noticeable quality degradation.
Thus, it is intuitive to reuse at the root cause, the channel level.
Moreover, different channel groups encode distinct spatio-temporal information. 
Reusing them at the token level inevitably loses certain information.
\Sect{sec:eval:quality} shows that our channel-level reuse preserves much higher quality against other token-level approaches~\cite{bolya2023token, yoo2024adaptiv}.

\para{Thresholds.}
% While our reuse technique can achieve greater compute savings, a critical 
The next question is how to determine the appropriate thresholds, $\theta_\text{th1}$ and $\theta_\text{th2}$.
Our key observation is that the threshold impact on the final output is insensitive to the input prompts, as shown in \Fig{fig:threshold_prompt}.
Thus, we can predetermine the threshold values offline and apply them to all inputs.

To further simplify hardware implementation, we restrict all thresholds to powers of 2.
In our design, we set $\theta_{\text{th1}}$ and $\theta_{\text{th2}}$ to preserve the most significant 5 bits and 3 bits, assuming an 8-bit fixed-point representation, respectively. 
Specifically, if the first 5 bits of $x_a$ and $x_b$ are the same (computed via a bitwise XOR), then $x_b$ fully reuses the partial results of $x_a$. 
If the first 3 bits of $x_a$ and $x_b$ are the same, the most significant 3 bits of the partial result for $x_b$ are reused from $x_a$, while only the subsequent 4 bits of $x_b$ are explicitly computed. 
\Sect{sec:eval:sens} further the impact of different thresholds on model performance and generative quality.
\section{Architectural Design}
\label{sec:arch}

Although our channel-wise reuse algorithm significantly reduces the total number of operations across all computational blocks, its reuse-based data access patterns are inherently incompatible with off-the-shelf accelerators, e.g., a systolic array.
To address this, we introduce a co-designed architecture that natively supports our algorithm. 
We first present the overall architectural design in \Sect{sec:arch:ov}, and then explain two key components: a PE array with a tailored dataflow in \Sect{sec:arch:dataflow}, and a dedicated data dispatcher in \Sect{sec:arch:dispatcher} to guarantee high PE utilization during the computation.

\subsection{Overview}
\label{sec:arch:ov}

\begin{figure}
    \centering
    \includegraphics[width=\columnwidth]{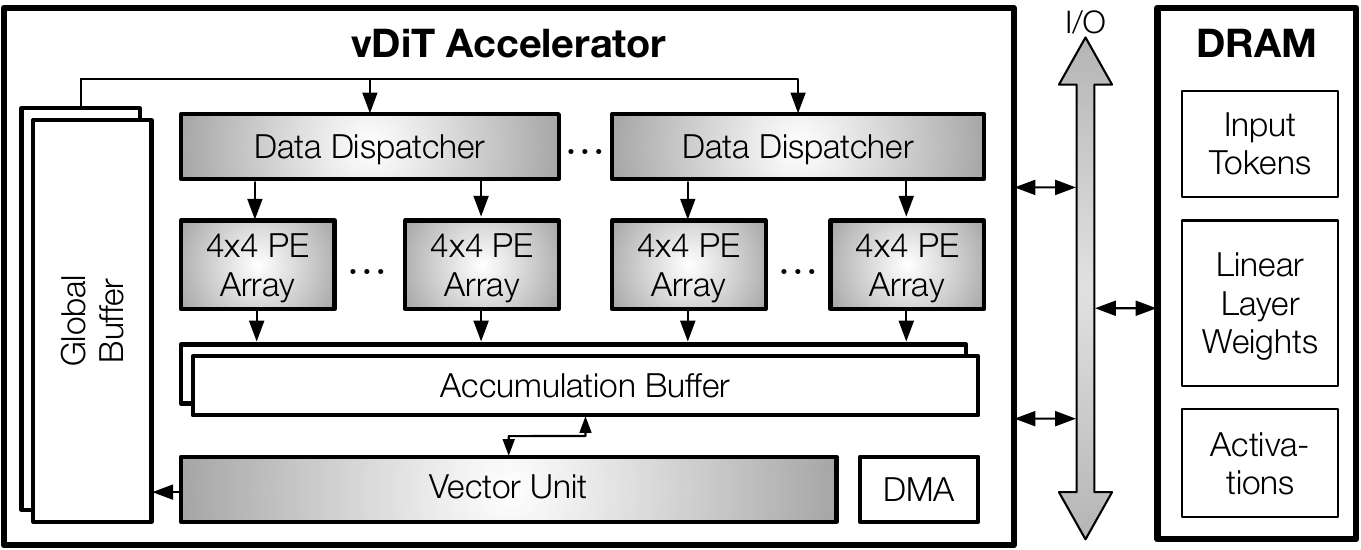}
    \caption{The overall architecture design, which comprises a set of data dispatchers, multiple PE arrays, and a vector unit.}
    \label{fig:arch}
\end{figure}

\para{Motivation.}
\hl{As shown in \mbox{\Fig{fig:idea}}, the proposed algorithm has various reuse patterns, which make it difficult to exploit efficiently using only GPU kernels due to its fine-grained, data-dependent execution.
Directly implementing our idea on a GPU would lead to severe warp divergence, irregular memory accesses, and low tensor core utilization.
These challenges motivate us to propose a dedicated architectural support for our algorithm.
}

\para{Overview.}
\Fig{fig:arch} illustrates the overall architecture of our vDiT accelerator. 
Our design comprises a set of PE arrays and a vector unit.
A subset of PE arrays is also coupled with a data dispatcher.
The PE arrays are responsible for compute-intensive matrix–matrix operations in vDiTs, e.g., self-attention, while the vector unit handles element-wise and vector operations, including Softmax. 
Each PE array is built upon a classic systolic array design with $4 \times 4$ PEs and is augmented with additional support to enable our reuse-aware datapath, as shown in \Fig{fig:pe_design}. 
The data dispatchers are used to feed input data into the PE arrays; their scheduling algorithm is explained in \Sect{sec:arch:dispatcher}.
Meanwhile, to support the pipelining between computation and data fetching, both the global buffer and the accumulation buffer are designed to be double-buffered.

\subsection{PE Array}
\label{sec:arch:dataflow}

\begin{figure}[t]
\centering
\begin{minipage}[t]{0.54\columnwidth}
  \centering
  \includegraphics[width=\columnwidth]{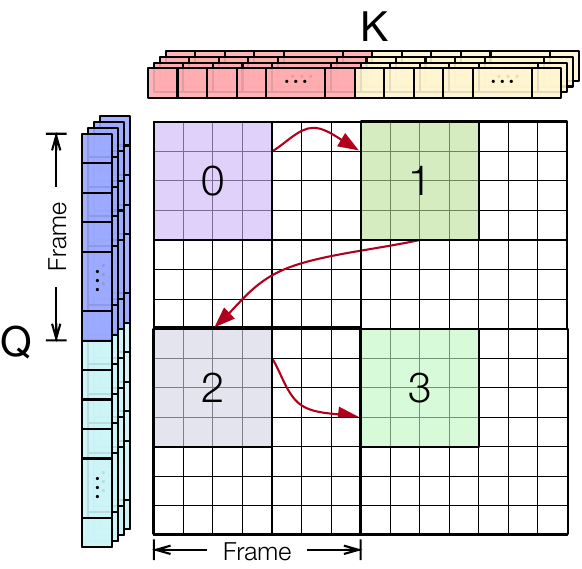}
  \caption{The computation order of self-attention computation when reusing along the t-axis. Each PE array is responsible for a group of tiles, as highlighted in colors. The numbers denote the computation order.}
  \label{fig:dataflow}
\end{minipage}
\hspace{2pt}
\begin{minipage}[t]{0.43\columnwidth}
  \centering
  \includegraphics[width=\columnwidth]{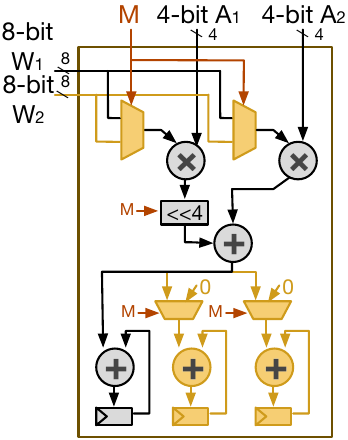}
  \caption{The design of one PE unit, which is built upon a canonical 8-bit MAC unit. The yellow parts highlight our augmented logics.}
  \label{fig:pe_design}
\end{minipage}
\end{figure}

\begin{figure*}
    \centering
    \includegraphics[width=\linewidth]{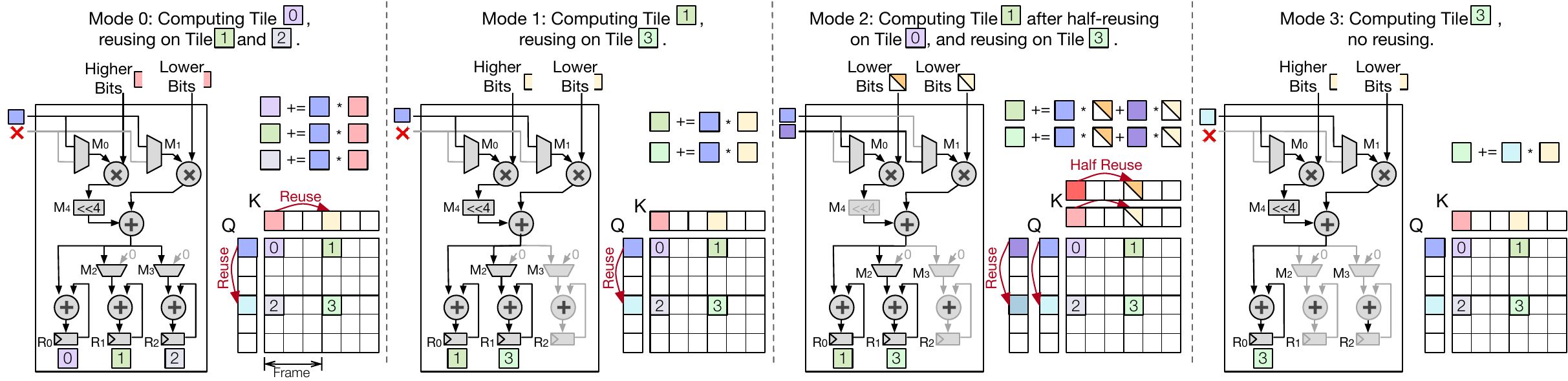}
    \caption{An overview of our reuse-aware dataflow for a single PE. In this example, 3 blocks in $Q$ and $K$ stand for one frame, i.e., $3 \times 3$ tiles are a frame in the attention map. We show four representative cases to show the flexibility of our PE. The control logics are omitted in PEs. The gray color indicates the parts that are disabled during the computation.}
    \label{fig:pe_mode}
\end{figure*}

We first illustrate the overall computation order of our reuse algorithm.
We then describe our augmentation to an 8-bit fixed-point multiply-and-accumulate (MAC) unit.
Lastly, we show how the PE is configured to support various data computation patterns.

\para{Computational Order.}
We first describe our computation order, which minimizes on-chip data movement, and how the workload is distributed across multiple PE arrays. 
Here, we use self-attention as an example.
As shown in \Fig{fig:dataflow}, we organize every $2 \times 2$ frames in the attention map as a group and process them together.
Specifically, we map the same $N \times N$ tile across the $2 \times 2$ frames to a single PE array.
In \Fig{fig:dataflow}, $N$ is 4.
Here, we compute the partial accumulations on a channel group basis, since the reuse direction is the same within one channel group.
For instance, when the reuse direction is along the t-dimension, the computational order is shown in \Fig{fig:dataflow}.
We first process the $4 \times 4$ attention elements in frame 0, and then we sequentially compute the same elements for frames 1, 2, and 3.
Other reuse directions can be done similarly.
Once we finish computing the partial attention scores within a channel group, we store the intermediate results in the accumulation buffer.

\para{PE Design.}
\Fig{fig:pe_design} shows our augmentation of a canonical 8-bit fixed-point MAC unit, with the added logic highlighted in yellow.
Specifically, we add two additional registers in each PE to store more intermediate results, along with extra multiplexers to select data from different sources.
The control signal, denoted as \textit{M}, configures the PE to operate in different execution modes. 
The following paragraphs explain how these configurations enable PEs to support various data reuse patterns.

\para{Reuse-Aware Dataflow.}
Based on our channel-wise reuse algorithm, reuse patterns vary across different frames within a $2 \times 2$ group.
In \Fig{fig:pe_mode}, each frame has $3 \times 3$ tiles in the attention map.
To minimize the data traffic between PEs and the accumulation buffer, each PE is configurable to support different reuse patterns.
Our overall dataflow adopts an output-stationary pattern in a classic systolic array, with some changes to accommodate reuse.
We showcase four execution modes using reuse along the t-dimension as an example; reuse along the other two dimensions works similarly.
Additional modes are the variants of the four modes.

\underline{Mode 0.} 
This mode is used when computing tiles such as Tile 0, as annotated in \Fig{fig:pe_mode}.
When processing Tile 0, the partial attention score could be reused by tiles in subsequent frames, i.e., Tile 1 and Tile 2. 
The first example in \Fig{fig:pe_mode} shows this scenario, where the blue element in $Q$ and the pink element in $K$ are reused.
Three registers in PE are used to store the accumulated partial attention scores corresponding to the elements in Tile 0, Tile 1, and Tile 2.

During computation, the input element of K is read from the column input port, while the input element of Q is read through the first row input port; the second row input port is disabled.
After the partial attention score is computed, the result is accumulated into three corresponding registers.
The multiplexers $M_{2}$ and $M_{3}$ control whether the partial score should be accumulated into registers $R_{1}$ and $R_{2}$, respectively, based on the reuse mask described in \Sect{sec:algo:idea}.
After computing all channels in one channel group, the current accumulated attention score of Tile 0 is temporarily read out from all $R_{0}$ of the PE array to the accumulation buffer.

\underline{Mode 1.}
The second mode is used when computing tiles such as Tile 1 or Tile 2.
When reuse is applied along the t-dimension, elements in Tile 1 or Tile 2 can only be reused by Tile 3, as shown by the second case in \Fig{fig:pe_mode}.
In this example, we show how elements in $Q$ are reused, i.e., Tile 3 reuses the values in Tile 1.

In this mode, the partially accumulated attention score of Tile~2 is temporarily stored in $R_2$ to avoid reloading it from the accumulation buffer.
Registers $R_0$ and $R_1$ hold the accumulated partial sums for Tile 1 and Tile 3, respectively.
During computation, the input values of $Q$ and $K$ are read from the row and column input ports, respectively.
The resulting partial sum is accumulated into $R_0$ for Tile 1.
Meanwhile, it is also selectively accumulated into $R_1$ based on the reuse criterion.
Note that, if the resulting partial sum is considered ``\textit{half-reuse}'', i.e., the leading bits are reused, the multiplexer $M_2$ is configured to accumulate only the leading $h$ bits into $R_{1}$.

\underline{Mode 2.}
The third mode is also used when computing tiles such as Tile 1 or Tile 2.
However, in this case, Tile 1 and Tile 2 have already partially reused the results from Tile 0, i.e., the leading $h$-bit partial sums.
Thus, Tile 1 and Tile 2 still need to calculate the remaining partial sum of the subsequent 4 bits after the $h$th bit.

Specifically, in the third example in \Fig{fig:pe_mode}, the values in $Q$ are reused, while the values in $K$ are partially reused.
To use our PE, the lower bits of two $K$ elements are fed into two 4-bit column input ports, while two distinct 8-bit $Q$ values are read from the two row input ports.
The control logic configures the datapath such that the two $Q$ values are processed by separate $4 \times 8$ MAC units. 
The resulting products are then directly added without performing the shift operation. 
The accumulated result is stored in registers $R_0$ and $R_1$.
Note that, if reuse is not applied to $Q$, the corresponding multiplexer can simply be configured to not accumulate into $R_1$.

\underline{Mode 3.} 
The final mode is used when computing tiles in the last tile in a $2 \times 2$ group.
In this mode, input values from $Q$ and $K$ are fed into the PE through the row and column input ports, respectively.
$R_1$ and $R_2$ are always disabled, as no need to reuse elements from Tile 3, as shown in the last example of \Fig{fig:pe_mode}.
The computed partial sum is then accumulated into register $R_0$, which stores the accumulated results of Tile 3.

Note that, the datapath of the last example shows accumulation for a single full 8-bit multiplication result.
When values in K are partially reused, the PE can be configured, similar to \underline{Mode 2}.
Two lower-bit $K$ values are fed into two 4-bit column ports, so that we can compute two partial sums simultaneously. 
The only difference is that multiplexer $M_2$ is configured to always disable the accumulation of the partial sum into register $R_1$.

\para{Other Operations.}
Self-attention has the most complex reuse patterns, whereas other operations are simpler because reuse applies to only one input.
As a result, our PE can also accommodate reuse patterns in other vDiT operations, such as MLPs. 
Configuring the PE to support these operations is straightforward: register $R_0$ is used to compute the current value, while the remaining two registers are used to accumulate partial results when reuse is applied.
The overall configuration is similar to \underline{Mode 1}.
Meanwhile, we also support ``half-reuse'' in those operations.
The configuration is similar to \underline{Mode 2} in \Fig{fig:pe_mode}.

\subsection{Data Dispatcher}
\label{sec:arch:dispatcher}

\begin{figure}[t]
    \centering
    \subfloat[Low PE utilization schedule.]{
        \includegraphics[width=0.46\columnwidth]{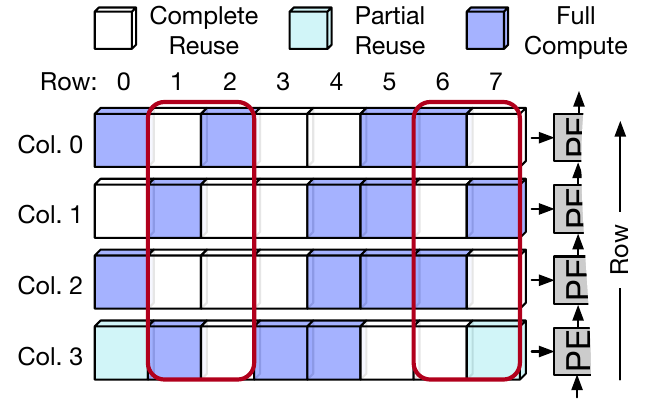}
        \label{fig:low_util}
    }
    \vspace{1pt}
    \subfloat[High PE utilization schedule.]{
        \includegraphics[width=0.46\columnwidth]{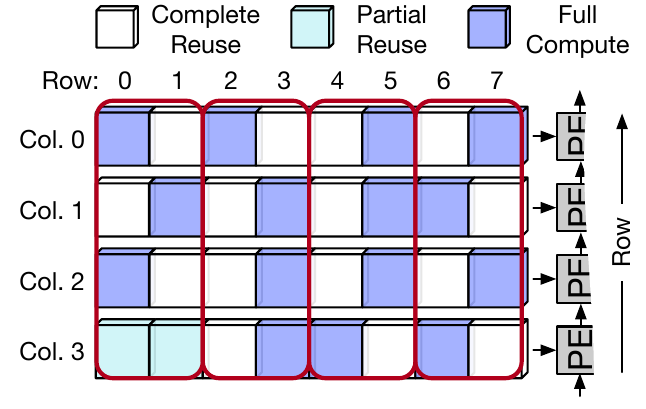}
        \label{fig:high_util}
    }
    \caption{Examples of two scheduling cases with identical data sparsity. The left example shows low PE utilization because many PEs perform no effective computation. In contrast, the right example achieves high PE utilization because two adjacent inputs can be merged and processed together. The red blocks highlight the rows that are merged.}
    \label{fig:schedules}
\end{figure}

\begin{figure*}
    \centering
    \includegraphics[width=\linewidth]{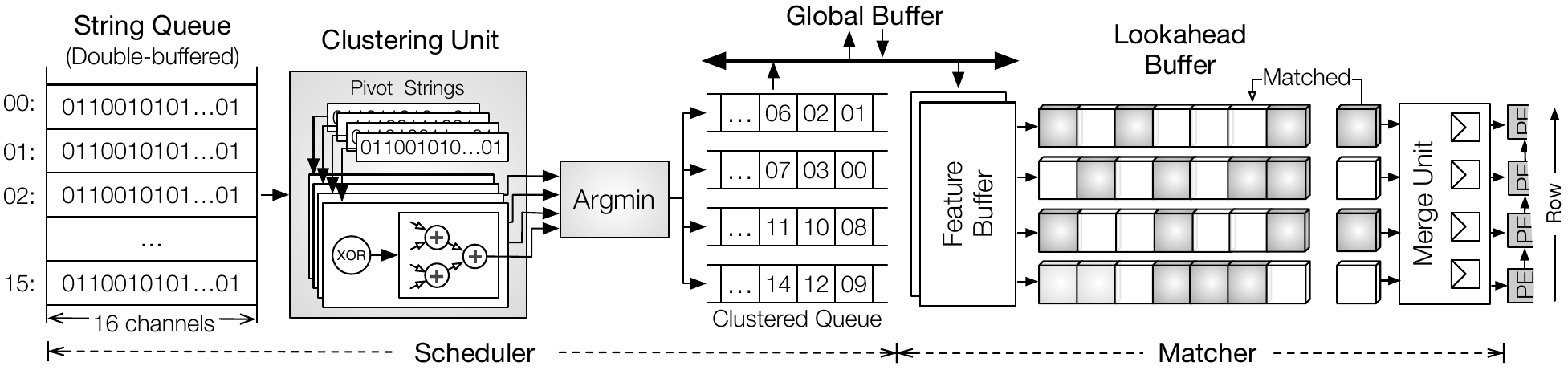}
    \caption{An overview of our data dispatcher design, which consists of a scheduler and a matcher. The scheduler groups tokens with similar reuse patterns together to improve the PE utilization. The matcher then pairs channels within each token group to combine compatible input pairs.}
    \label{fig:dispatcher}
\end{figure*}

\para{Issues.}
With our reconfigurable PE array design, we are able to support various reuse data access patterns.
However, given the fixed dataflow of the PE array, i.e., an output-stationary dataflow, sparse and irregular input patterns often lead to low PE utilization.

\Fig{fig:low_util} illustrates such a case. 
Here, we show only one side of the input data, i.e., keys $K$, while assuming that the other input, queries $Q$, is fully dense. 
Note that, $K$ is streamed from the column ports of the PE array.
Each column port of the PE array accepts 8-bit data, whereas each row port can read at most 16-bit data, corresponding to two input elements.
In principle, if two rows of $K$ are sufficiently sparse, they could be merged and processed at once to improve PE utilization.
However, the sparsity pattern of $K$ is highly irregular, as shown in \Fig{fig:low_util}.
The adjacent two rows rarely exactly match and allow merging; thus, many PEs remain idle since fully reused elements skip computation entirely.

In contrast, an ideal case is shown in \Fig{fig:high_util}, where every pair of adjacent rows can be merged. 
For instance, the first and second rows can be combined and fed into the PE array within a single cycle.
Note that, the only difference between \Fig{fig:low_util} and \Fig{fig:high_util} is the computation order.
Thus,  in this subsection, we design a data dispatcher to reorder computations and improve PE utilization.

\para{Design.}
\Fig{fig:dispatcher} illustrates the overall design of our data dispatcher, which consists of a scheduler and a matcher.
The scheduler groups tokens such that those with similar reuse patterns are dispatched to the PE array together. 
The matcher then further pairs channels within each token group to combine compatible input pairs, similar to the example in \Fig{fig:high_util}, further improving PE utilization.
The following paragraphs explain these two components.

The left side of \Fig{fig:dispatcher} illustrates our scheduler. 
The reuse pattern of each token is represented as a binary string, where ``0'' denotes full reuse and ``1'' denotes half-reuse or no reuse.
These token strings are stored in a string queue with 64 entries.
Each entry contains a 16-bit string, i.e., 16 channels are processed as a batch.

To cluster these 64 strings into four groups, the clustering unit first randomly picks four strings as pivot strings.
It then computes the Hamming distance between each input string and each pivot string, and uses an adder tree to accumulate the total differences.
Based on the results, each input string is assigned to the cluster corresponding to the pivot string with the minimum distance and is enqueued into the corresponding clustered queue.
Note that, each clustered queue has a fixed maximum capacity, i.e., 16 strings in this case. 
If the preferred queue is already full, the input string is assigned to the queue with the next smallest distance instead. 
Through this process, each clustered queue ultimately contains exactly 16 input strings for subsequent computing.

Once a batch of input strings is clustered, each cluster is mapped to one PE array.
We design one matcher for one PE array.
Each matcher reads the corresponding token features from the global buffer into the feature buffer. 
In each cycle, the merge unit examines one channel value from each of the four input tokens and checks whether these values can be merged with any other inputs. 
If a match is found, the merge unit combines two input values and feeds them into the PE array.
Otherwise, the single input values are fed into the PE array.
Note that, a scheduler is required for both sides of the data, i.e., $Q$ and $K$. 
However, only one side requires a matcher.
% In our design, it is $K$.
The other side follows the same merging policy implicitly.

\section{Experimental Setup}
\label{sec:exp}

\para{Experimental Methodology.}
We develop validated RTL implementations for the \proj hardware. The hardware is based on a systolic array architecture, consisting of $128 \times 128$ PE arrays, each with $4 \times 4$ PEs clocked at 1~GHz.
The PEs are designed for 8-bit fixed-point representation.
The hardware also has a scalar unit, which consists of 512 parallel lanes, each capable of performing the Softmax and other activation functions.
Both the global buffer and the accumulation buffer are implemented by SRAM.
The global buffer is 64~MB to store input data and model weights.
The Accumulation buffer is 32~MB in size to store output data.
Both buffers are double-buffered.
While we primarily evaluate \proj using this configuration, we will later show the sensitivity of \proj performance to different hardware resource configurations.

\para{Simulation Methodology.}
The RTL is synthesized using Synopsys tools and laid out using Cadence tools in TSMC 16nm FinFET technology, with SRAMs generated by an ARM compiler. 
Power is simulated using Synopsys PrimeTimePX, with full annotated switching activity. 
The off-chip DRAM is modeled as 16 DDR5-6400 channels based on Micron specifications~\cite{micronddr5}, and DRAM energy is estimated using Micron’s System Power Calculator~\cite{microdrampower}. 
We build a cycle-level simulator of the architecture with the latency and energy of each component parameterized from post-synthesis results.

\begin{table} 
\caption{\hl{Area breakdown of \mbox{\proj} architecture}.}
\resizebox{0.95\columnwidth}{!}{
\renewcommand*{\arraystretch}{1}
\renewcommand*{\tabcolsep}{5pt}
\begin{tabular}{ ccccc } 
\toprule[0.15em]
  & \textbf{$4 \times 4$ PE Array} &  \textbf{Dispatcher} & \textbf{On-Chip Buffer} & \textbf{Total} \\ 
\midrule[0.05em]
Configuration & $128 \times 128$ & $64 \times 32$ & 64~MB+32~MB & -- \\
Area (mm$^2$) & 25.1 & 1.3 & 68.4 & 94.8 \\
\bottomrule[0.15em]
\end{tabular}
}
\label{tab:area}
\end{table}

\para{Area.} \Tbl{tab:area} shows that the overall area of \proj is 94.8~mm$^2$ at 16~nm.
Compared to a baseline systolic array with 96~MB on-chip buffer, our augmentation on PE arrays and our data dispatcher introduce area overheads of 6.6\% and 1.3\%, respectively.

\para{Models and Metrics.}
\hl{We evaluate \mbox{\proj} on four widely adopted vDiT models: HunyuanVideo~\mbox{\cite{kong2024hunyuanvideo}}, Wan~\mbox{\cite{wan2025}}, CogVideoX~\mbox{\cite{yang2024cogvideox}} and TurboDiffusion~\mbox{\cite{zhang2025turbodiffusion}}.
Without further notice, we generate 5-second videos with $480\times540$ resolution. 
Note that, TurboDiffusion is a distilled model with only 4 timesteps.
To evaluate the generative quality, we use the VBench~\mbox{\cite{huang2024vbench}} as the video quality metric, which consists of 950 benchmark prompts.
% In our experiments, we generate 5 videos for each of the 950 benchmark prompts using different random seeds and then average the VBench scores.
The generated videos are evaluated across 16 aspects from VBench.}
For image quality evaluation, we use PSNR, SSIM, and LPIPS, and compare against the videos generated by the baseline models.

\para{Hardware Baselines.}
We compare six hardware baselines:
\begin{itemize}
    \item \mode{A100}: a Nvidia A100 GPU with 80~GB HBM~\cite{a100}.
    \item \mode{H100}: a Nvidia H100 GPU with 80~GB HBM~\cite{h100}.
    \item \mode{Cambricon-D}~\cite{kong2024cambricon}: approximates the computation between adjacent timesteps to accelerate the diffusion process.
    \item \mode{AdapTiV}~\cite{yoo2024adaptiv}: leverages the value similarity among tokens to reduce the effective computation for image classification.
    \item \mode{Exion}~\cite{heo2025exion}: exploits inter- and intra-timstep sparsity among intermediate results to skip computation. 
    \item \mode{Ditto}~\cite{kim2025ditto}: also leverages the value similarity between adjacent timesteps to reduce the effective computation.
\end{itemize}
All accelerator baselines are implemented using 8-bit fixed-point precision and scaled to the same number of PEs, $512 \times 512$, at 1~GHz.
The hardware simulation is modeled under 16~nm technology.
The energy and performance of \mode{A100} and \mode{H100} are measured using the built-in power sensing circuitry on Nvidia A100 and H100.

\para{Software Baselines.}
\hl{In addition to six hardware baselines, we also compare \mbox{\proj} against four algorithmic baselines.
Specifically, we compare one block caching technique, \mbox{\mode{PAB}}~\mbox{\cite{zhao2024real}}.
We also compared three sparse attention acceleration methods, including \mbox{\mode{MInference}~\cite{jiang2024minference}}, \mbox{\mode{SVG}~\cite{xi2025sparse}} and \mbox{\mode{SVG2}~\cite{yang2025sparse}}.
The performance numbers of all software baselines are measured on a Nvidia A100 GPU.
}

\para{Variants.}
\hl{
We evaluate three variants of \mbox{\proj} to separate the contributions in our paper:} 
\begin{itemize}
    \item \hl{\mbox{\mode{\proj-pe}}: a variant which only consists of our modified PE arrays without our data dispatcher.}
    \item \hl{\mbox{\mode{\proj-pe-sh}}: a variant which only consists of our modified PE arrays and the scheduler in our data dispatcher, but without the matcher.}
    \item \hl{\mbox{\mode{\proj-pe-m}}: a variant which only consists of our modified PE arrays and the matcher in our data dispatcher, but without the scheduler.}
    \item \hl{\mbox{\mode{\proj}}: our full-fledged design with all optimizations.}
\end{itemize}

For both variants, we set $\theta_\text{th1}$ and $\theta_\text{th2}$ to preserve the first 5 bits and first 3 bits of accuracy, respectively. 
We do not show a GPU implementation of our algorithm because its reuse-based data access patterns are fundamentally incompatible with existing attention acceleration frameworks~\mbox{\cite{dao2022flashattention, kwon2023efficient}}, e.g., FlashAttention.
Thus, a direct comparison of GPU implementations is not meaningful.
\section{Evaluation}
\label{sec:eval}

\subsection{Generative Quality}
\label{sec:eval:quality}

\begin{figure}[t]
\centering
\subfloat[\hl{PSNR. Higher is better.}]{
    \label{fig:psnr}
    \includegraphics[width=\columnwidth]{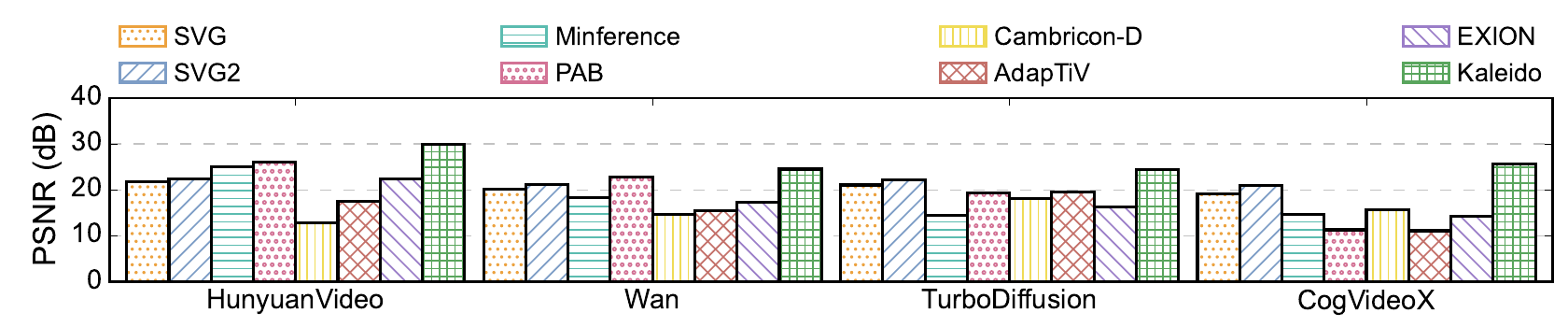}}
\\
\subfloat[\hl{SSIM. Higher is better.}]{
    \label{fig:ssim}
    \includegraphics[width=\columnwidth]{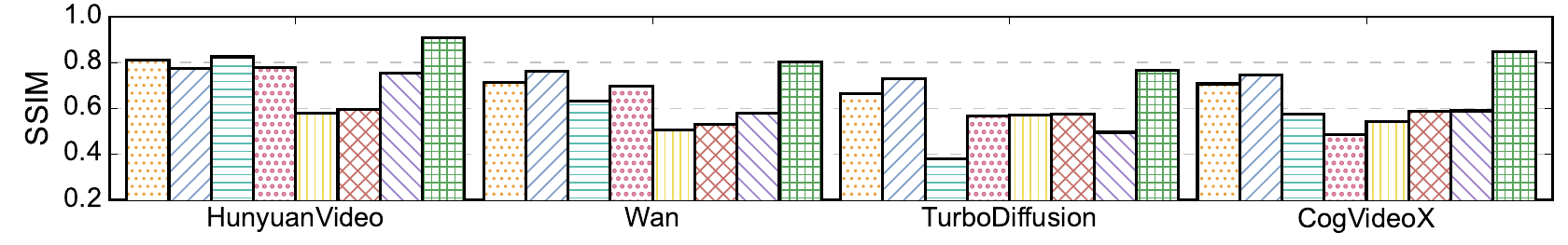}}
\\
\subfloat[\hl{LPIPS. Lower is better.}]{
    \label{fig:lpips}
    \includegraphics[width=\columnwidth]{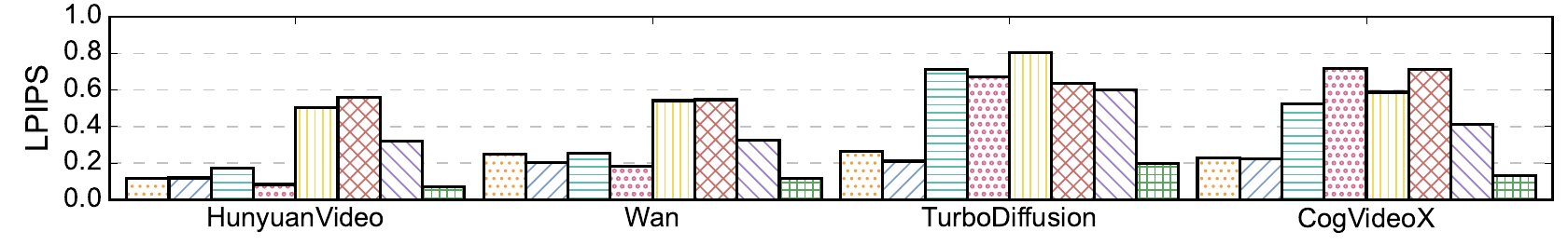}}
\\
\subfloat[\hl{VBench score. Higher is better.}]{
    \label{fig:vbench}
    \includegraphics[width=\columnwidth]{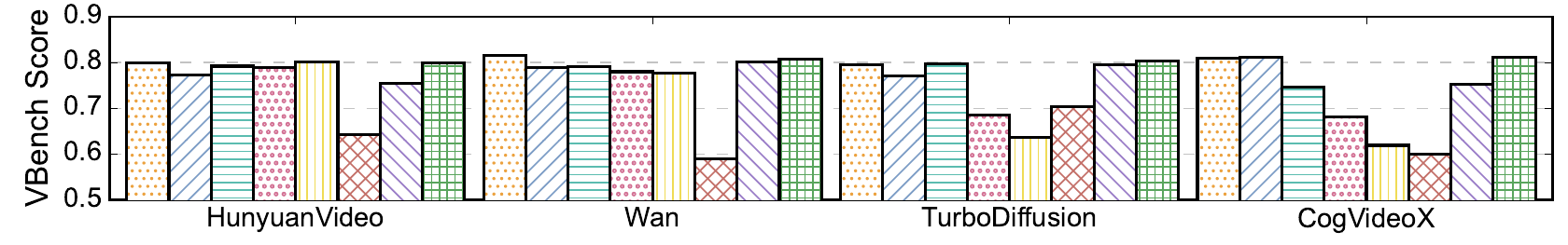}}
\\
\subfloat[Qualitative comparison on HunyuanVideo. Please zoom in to check details.]{
    \label{fig:qual_cmp}
    \includegraphics[width=\columnwidth]{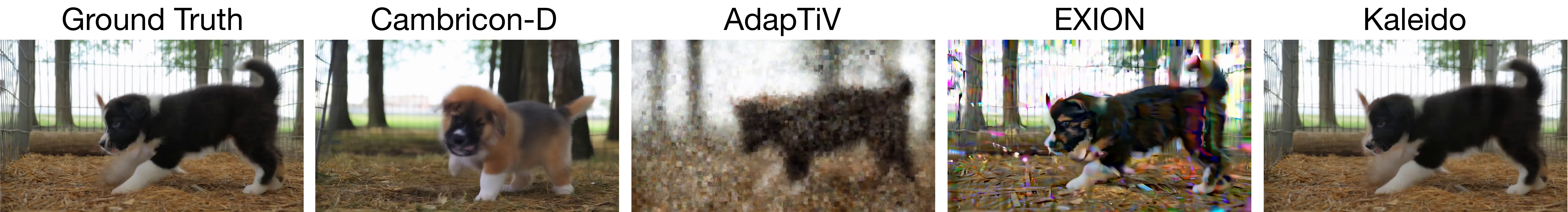}}
\caption{\hl{Video generation quality comparison.} 
% All subfigures share the same legend.
}
\label{fig:acc}
\end{figure}

Here, we primarily compare against three approximation baselines: \mode{Cambricon-D}, \mode{AdapTiV}, and \mode{Exion}.
\hl{We also include four software-optimized baselines here.}
Since the algorithm proposed in \mode{Ditto} is lossless, we do not include it in the quality comparison.
% The detailed configurations are shown in \Tbl{??}.

\hl{\mbox{\Fig{fig:acc}} compares the video generation quality of \mbox{\proj} with the three approximation methods shown above.
% We test across three mainstream vDiT models: HunyuanVideo~\cite{kong2024hunyuanvideo}, Wan~\cite{wan2025}, and TurboDiffusion~\cite{zhang2025turbodiffusion}.
Across all four metrics, \mbox{\mode{\proj}} consistently delivers the best quality.
For instance, \mbox{\mode{\proj}} improves reconstruction fidelity by a large margin in terms of PSNR.
\mbox{\mode{\proj}} achieves 29.9~dB, 24.5~dB, 24.4~dB, and 26.5~dB on HunyuanVideo, Wan, TurboDiffusion, and CogVideoX, respectively.
% The PSNRs of other methods are all below 20~dB.
In particular, \mbox{\mode{\proj}} achieves 17.0~dB over \mbox{\mode{Cambricon-D}} on HunyuanVideo.
Across all models, \mbox{\mode{\proj}} achieves over 6~dB higher than all prior accelerator methods.
Even compared with the software-optimized baselines, \mbox{\mode{\proj}} consistently achieves the highest results over the image-based metrics.}
This shows that the channel-wise reuse algorithm is the key to preserving high pixel-level precision. 
Similar trends can be seen in SSIM (\mbox{\Fig{fig:ssim}}) and LPIPS (\mbox{\Fig{fig:lpips}}).
For instance, \mbox{\mode{\proj}} achieves 0.87 to 0.90 in SSIM, while others only achieve 0.50 to 0.78 in SSIM.

% Finally, on the VBench metric (\Fig{fig:vbench}), which captures high-level video quality and temporal coherence, \proj maintains scores around 0.81–0.82, exceeding the best prior methods. 
Finally, on the VBench metric (\Fig{fig:vbench}), which captures high-level video quality, \proj maintains the same scores as the baseline algorithms (around 0.81), exceeding the best prior methods. 
\Fig{fig:qual_cmp} further showcases the qualitative results of all methods.
Visually, \mode{\proj} preserves much higher quality compared to others.

\subsection{Performance Comparison}
\label{sec:eval:perf}

\begin{figure}[t]
\centering
\subfloat[\hl{Speedup evaluation. Higher is better.}]{
    \label{fig:speedup}
    \includegraphics[width=\columnwidth]{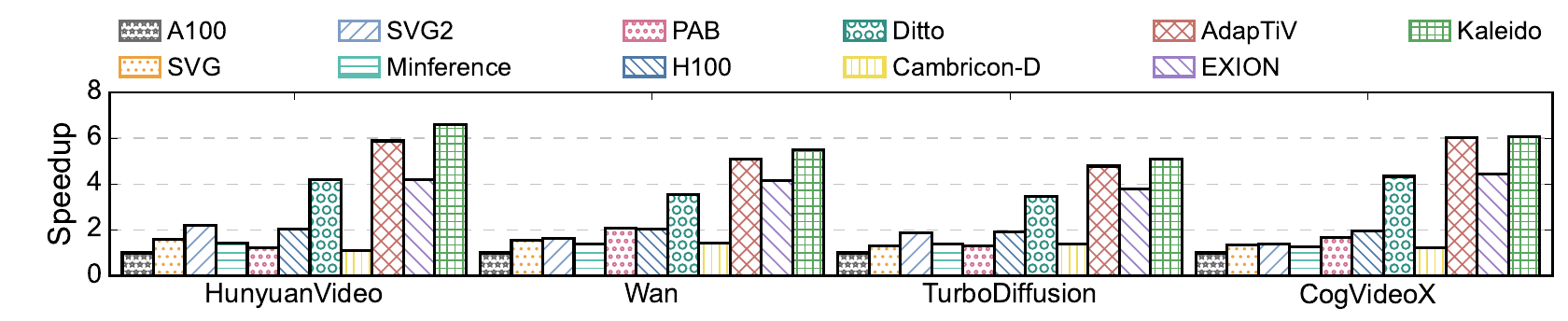}}
\\
\subfloat[\hl{Energy savings. Higher is better.}]{
    \label{fig:energy}
    \includegraphics[width=\columnwidth]{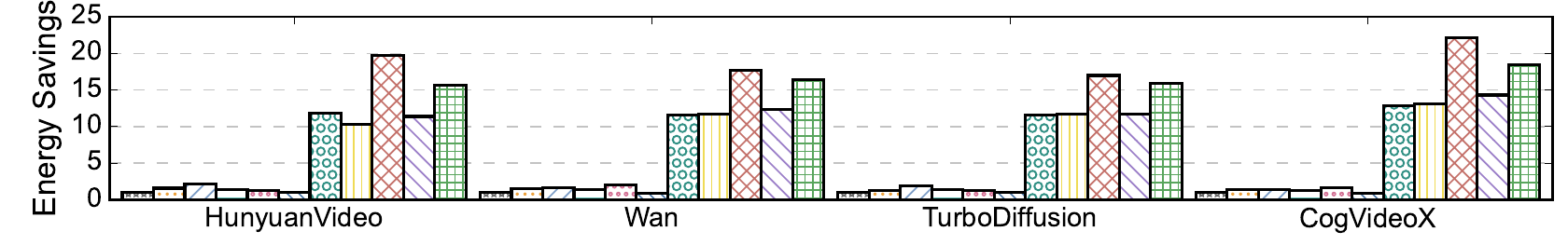}}
\\
\subfloat[Overall operation reduction. Lower is better.]{
    \label{fig:flops}
    \includegraphics[width=\columnwidth]{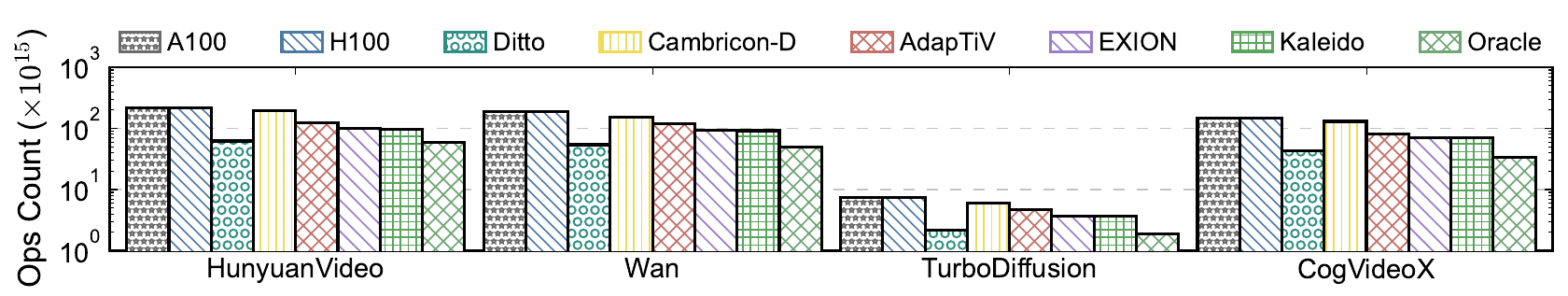}}
\caption{\hl{Overall performance evaluation against prior work. }
% All subfigures share the same legend.
}
\label{fig:perf}
\end{figure}

\para{Performance.}
\hl{\mbox{\Fig{fig:speedup}} compares the end-to-end performance comparison of different accelerators.
Here, we compare \mbox{\mode{\proj}} against four accelerators: \mbox{\mode{Cambricon-D}}, \mbox{\mode{AdapTiV}}, \mbox{\mode{Exion}} and \mbox{\mode{Ditto}}.
Meanwhile, we also compare four GPU-based optimizations: \mbox{\mode{SVG}}, \mbox{\mode{SVG2}}, \mbox{\mode{Minference}}, \mbox{\mode{PAB}}, 
Across all four vDiT models, \mbox{\mode{\proj}} achieves the highest speedup.
Specifically, \mbox{\mode{\proj}} delivers 6.6$\times$, 5.6$\times$, 5.2$\times$, and 6.1$\times$ speedup over \mbox{\mode{A100}} on HunyuanVideo, Wan, TurboDiffusion, and CogVideoX, respectively.}

While methods like \mode{AdapTiV} also achieve relatively high performance, over 5$\times$ speedup, \mode{AdapTiV} achieves much lower generative quality compared to \mode{\proj}.
Because its acceleration technique, token merging, is designed for
image classification and fundamentally ill-suited for generative tasks.
\mode{Cambricon-D} yields marginal speedups because its architecture is primarily designed for CNN-like models and cannot accelerate self-attention, which is the primary computation bottleneck in modern vDiT models.
Finally, both \mode{Exion} and \mode{Ditto} struggle to achieve high speedups due to the irregularity of sparse computation in video generation, especially for long token sequences.
Both designs have to deal with the workload imbalance between PEs.
In contrast, \mode{\proj} co-designs a data dispatcher to tame the workload imbalance and improve the PE utilization.
\Sect{sec:eval:abl} further dissects the impact of data dispatch.

\para{Energy.}
\hl{
\mbox{\Fig{fig:energy}} shows the energy savings of all evaluated hardware baselines normalized to \mbox{\mode{A100}}.
Here, energy saving is defined as the ratio between the energy consumption of the baseline, i.e., \mbox{\mode{A100}}, and that of the corresponding accelerator design.
Overall, \mbox{\mode{\proj}} achieves 14.8$\times$, 15.6$\times$, 15.2$\times$, and 18.4$\times$ energy savings on HunyuanVideo, Wan, TurboDiffusion, and CogVideoX, respectively.
The latency overhead of the data dispatcher is negligible.
Note that, the scheduler in the data dispatcher can be pipelined and overlapped with the subsequent matrix computation.
}

\hl{While \mbox{\mode{AdapTiV}} exhibits higher energy savings than \mbox{\mode{\proj}}, its energy savings primarily come from its aggressive merge ratio.
By merging tokens, \mbox{\mode{AdapTiV}} can directly reduce the overall token sequence length and reduce a large amount of off-chip DRAM accesses.
However, any sparsity computation techniques, e.g., \mbox{\mode{\proj}} or \mbox{\mode{Exion}}, still have to maintain the full token sequence and cannot eliminate the off-chip data traffic.
Nevertheless, as we mentioned earlier, the token merging technique in \mbox{\mode{AdapTiV}} introduces unacceptable accuracy loss and is practically unacceptable.
Remaining baselines all achieve lower energy savings compared to \mbox{\mode{\proj}}.
Note that, the data dispatcher contributes 0.1\% of the total energy.}

\para{Operation Reduction.}
\Fig{fig:flops} shows the total operation count for each hardware baseline.
Here, we show two numbers of \proj: \mode{\proj} and \mode{Oracle}.
\mode{\proj} is the effective operation reduction achieved during actual execution under our hardware limitations.
\mode{Oracle}, on the other hand, is the ideal operation reduction without any hardware limitations.
For other hardware baselines, we show their operation reduction based on their algorithms without considering their hardware constraints.

We show that, by exploiting spatio-temporal correlations and reusing intermediate channel-wise results, \mode{Oracle} can achieve up to 85\% of operation reduction, eliminating a massive amount of computations.
However, with hardware constraints, the operation reduction decreases to around 60\%.
\hl{There are two main reasons leading to the gap between \mbox{\mode{Oracle}} and \mbox{\mode{\proj}}.
First is an algorithmic constraint. 
In \mbox{\mode{\proj}}, each channel group is restricted to reuse along a single fixed direction to simplify scheduling and hardware mapping.
In contrast, \mbox{\mode{Oracle}} allows each channel to reuse in any direction.
Second is hardware-aware mapping overhead.
Due to PE-array granularity, some operations marked as reusable may still be executed as part of a regular PE tile.}
% For instance, on HunyuanVideo, \mode{\proj} reduces the total operations by 54.3\%.
% However, \mode{Oracle} can further reduce the operations to 85.1\%.
\mode{Ditto} achieves the second-best theoretical operation reduction because it aggressively skips 0-value computations.
However, \mode{Ditto}'s adder tree design cannot efficiently process these irregular operations, leading to severe pipeline stalling and poor PE utilization.

\subsection{Ablation Study}
\label{sec:eval:abl}

\begin{figure}[t]
\centering
\includegraphics[width=\columnwidth]{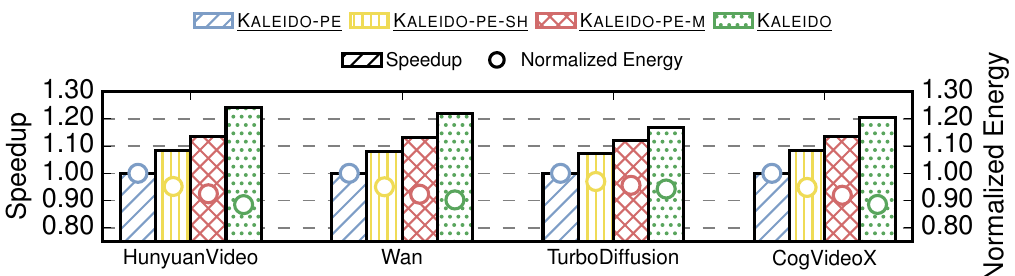}
\caption{\hl{Ablation study of the different hardware components in \mbox{\proj}.}}
\label{fig:abl_dispatcher}
\end{figure}

% \begin{figure}[t]
% \centering
% \begin{minipage}[t]{0.49\columnwidth}
%   \centering
%   \includegraphics[width=\columnwidth]{ablation_dispatcher}
%   \caption{\hl{Ablation study of \mbox{\proj} with and without data dispatcher. ``-S'': speedup, ``-E'': normalized energy.}}
%   \label{fig:abl_dispatcher}
% \end{minipage}
% \hfill
% \begin{minipage}[t]{0.49\columnwidth}
%   \centering
%   \includegraphics[width=\columnwidth]{sens_cs_speedup}
%   \caption{Sensitivity of performance to per-array PE size. The numbers are normalized to the default configuration with 4$\times$4 PEs per PE array.}
%   \label{fig:sens_chunk}
% \end{minipage}
% \end{figure}

\begin{figure}[t]
\centering
\begin{minipage}[t]{0.49\columnwidth}
  \centering
  \includegraphics[width=\columnwidth]{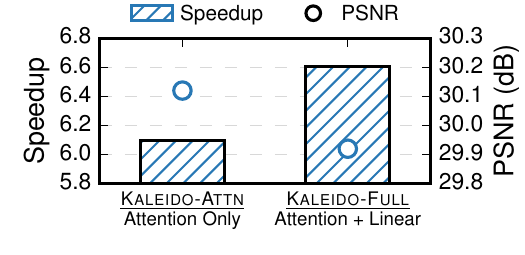}
  \caption{\hl{Ablation study of performance and quality on HunyuanVideo, with and without applying \mbox{\proj} to linear projection layers.}}
  \label{fig:abl_linear}
\end{minipage}
\hfill
\begin{minipage}[t]{0.49\columnwidth}
  \centering
  \includegraphics[width=\columnwidth]{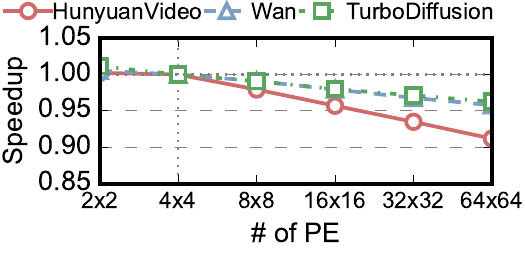}
  \caption{Sensitivity of performance to per-array PE size. The numbers are normalized to the default configuration with 4$\times$4 PEs per PE array.}
  \label{fig:sens_chunk}
\end{minipage}

\end{figure}

\para{Hardware Ablation.}
\Fig{fig:abl_dispatcher} shows an ablation study of the four variants in \mbox{\Sect{sec:exp}}: \mode{\proj-pe}, which includes only the reconfigurable PE array without the data dispatcher; \mode{\proj-pe-sh}, which adds only the scheduler in the data dispatcher; \mode{\proj-pe-m}, which adds only the matcher in the data dispatcher; and the full \mode{\proj} design.
Both speedup and energy are normalized to\mode{\proj-pe}.

Across three vDiT models, we show that introducing our data dispatcher achieves 1.2$\times$ speedup and 9.0\% of energy reduction.
This is because, without the dispatcher, our channel-wise reuse algorithm would introduce irregular sparsity in computation and lead to PE under-utilization.
\hl{The variant with the matcher achieves higher speedup than the variant with only the scheduler.}
With our data dispatcher, we improve the PE utilization from 69.4\% to 85.2\% by clustering tokens with similar reuse patterns and taming the irregularity during computation.
\hl{The scheduler in the data dispatcher is pipelined and overlapped with the subsequent computation.
The latency overhead of our data dispatcher is negligible.}

\para{Algorithmic Ablation.}
\hl{
\mbox{\Fig{fig:abl_linear}} compares the performance and generative quality of applying our algorithm either to attention layers only or to all layers.
We report the results on HunyuanVideo.
Compared with applying \mbox{\proj} only to attention layers, the full design achieves higher speedup by exploiting additional channel-wise reuse opportunities in the linear projection layers.
Meanwhile, the PSNR remains nearly unchanged.
It shows that extending reuse to linear layers preserves generation quality while providing additional acceleration.
}

\subsection{Sensitivity Study}
\label{sec:eval:sens}

\para{PE Size.}
\Fig{fig:sens_chunk} shows the sensitivity of speedup to the per-array PE size, where we vary the number of PEs per array while retaining the total number of PEs constant. 
All numbers are normalized to the default 4$\times$4 configuration. 
Overall, speedup degrades slightly as the per-array PE size increases.
This is because larger PE arrays are less flexible in handling irregular reuse patterns, thus reducing PE utilization. 
Nevertheless, \proj remains robust even with larger PE arrays. 
Meanwhile, further reducing the array size to 2$\times$2 achieves marginal benefit.
Thus, we choose the 4$\times$4 configuration.

\begin{figure}[t]
\centering
\subfloat[Speedup.]{
    \label{fig:sens_speedup}
    \includegraphics[width=0.49\columnwidth]{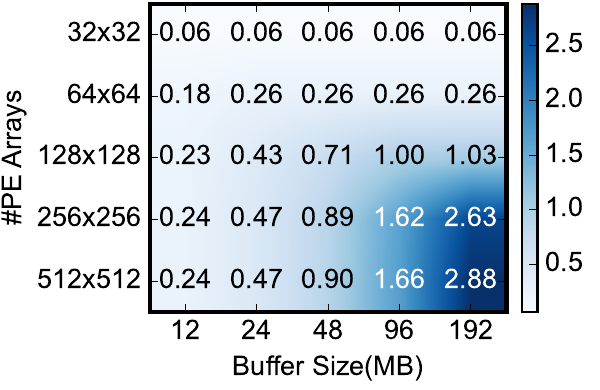}}
\hfill
\subfloat[Normalized energy.]{
    \label{fig:sens_energy}
    \includegraphics[width=0.49\columnwidth]{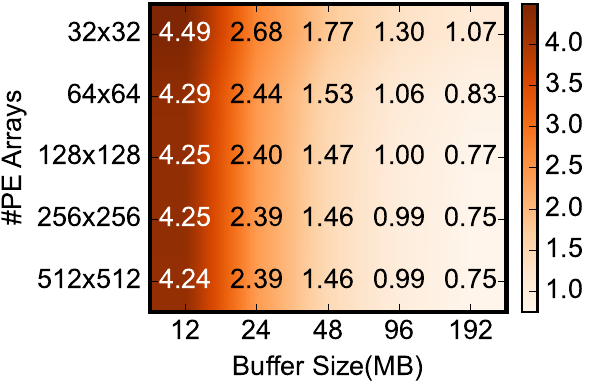}}
\caption{Sensitivity of performance and energy efficiency to the number of PE arrays and buffer size. Data are normalized to the configuration with $128 \times 128$ PE arrays and 96~MB buffer.}
\label{fig:sens_pe_buffer}
\end{figure}

\para{PE Array \& Buffer Size.}
\Fig{fig:sens_pe_buffer} shows the sensitivity of performance and energy efficiency to the number of PE arrays and buffer size. 
All results are normalized to our default configuration, i.e., 128$\times$128 PE arrays and a 96 MB buffer.
Each PE array consists of 4$\times$4 PEs.
As shown in \Fig{fig:sens_speedup}, increasing the number of PE arrays generally improves the overall performance.
However, if the on-chip buffer is not large enough, off-chip data communication would impact the performance.
Larger buffers reduce off-chip memory traffic, often leading to higher performance.

\Fig{fig:sens_pe_buffer} further shows the normalized energy consumption under different PE and buffer configurations.
Increasing the number of PE arrays slightly reduces overall energy because it reduces execution time.
Meanwhile, a larger buffer size can better improve energy efficiency because it minimizes off-chip data traffic overhead.
Overall, both \Fig{fig:sens_speedup} and \Fig{fig:sens_pe_buffer} show that the number of PE arrays and the total size of on-chip buffer need to be balanced to achieve optimal performance and energy efficiency.

\begin{figure}[t]
\centering
\begin{minipage}[t]{0.49\columnwidth}
  \centering
  \includegraphics[width=\columnwidth]{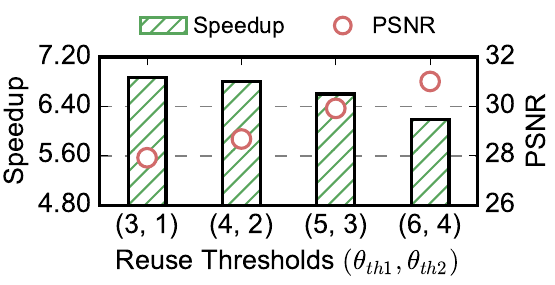}
  \caption{Sensitivity of speedup and generative quality to reuse thresholds, $\theta_\text{th1}$ and $\theta_\text{th2}$.}
  \label{fig:sens_thresholds}
\end{minipage}
\hfill
\begin{minipage}[t]{0.49\columnwidth}
  \centering
  \includegraphics[width=\columnwidth]{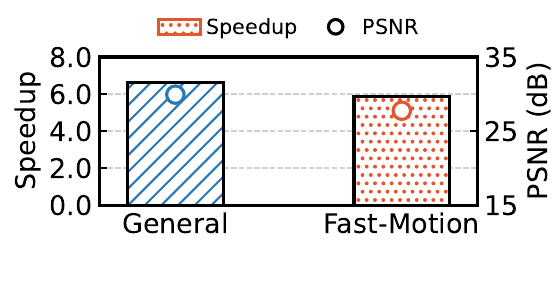}
  \caption{\hl{Comparison of performance and generative quality under fast-motion prompts.}}
  \label{fig:fast_motion}
\end{minipage}
\end{figure}

\para{Thresholds.}
In our algorithm design, our default threshold setting is to set $\theta_\text{th1}$ and $\theta_\text{th2}$ to preserve the most significant 5 bits and 3 bits of accuracy, respectively.
In \Fig{fig:sens_thresholds}, we show the sensitivity of performance and accuracy to different threshold combinations.
The results show that preserving fewer bits leads to noticeable degradation in visual quality, whereas preserving more bits slows inference. 
We find that setting $\theta_{\text{th1}}$ and $\theta_{\text{th2}}$ to 5 and 3 provides a good trade-off between performance and quality.

\para{Fast Motion.}
\hl{
\mbox{\Fig{fig:fast_motion}} compares the performance and generative quality of \mbox{\proj} under general prompts and fast-motion prompts on HunyuanVideo.
We extract the prompts with fast motions from VBench.
Compared with general prompts, fast-motion prompts achieve slightly lower speedup because rapid temporal changes reduce the amount of reusable spatio-temporal correlation.
However, \mbox{\proj} still maintains substantial acceleration while preserving similar PSNR.
It means that our channel-wise reuse remains effective even for videos with fast motions.
}

\begin{figure}[t]
\centering
\begin{minipage}[t]{0.49\columnwidth}
  \centering
  \includegraphics[width=\columnwidth]{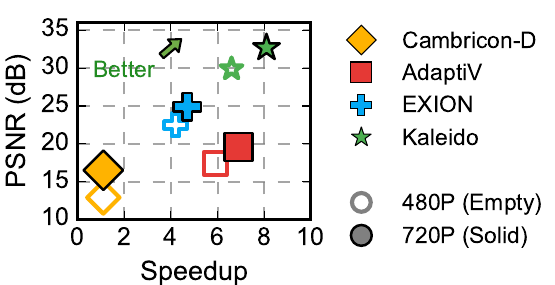}
  \caption{Scalability analysis on generating 720p high-resolution videos.}
  \label{fig:scalability}
\end{minipage}
\hfill
\begin{minipage}[t]{0.49\columnwidth}
  \centering
  \includegraphics[width=\columnwidth]{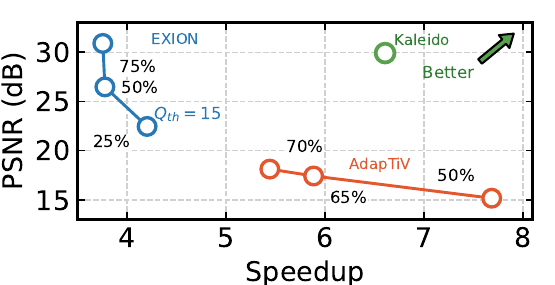}
  \caption{\hl{Comparison against \mbox{\mode{AdapTiV}} and \mbox{\mode{Exion}} under different configurations.}}
  \label{fig:adaptiv_exion}
\end{minipage}
\end{figure}

\subsection{Scalability Analysis}
\label{sec:eval:scale}

\Fig{fig:scalability} shows the scalability of different methods when generating higher-resolution (720p) videos.
As the resolution increases from 480p to 720p, \mode{\proj} consistently achieves both higher speedup and better visual quality compared to prior approaches. 
This is because our approach inherently exploits the spatio-temporal correlations across frames in the latent space.
These correlations are stronger as the resolution increases.
As a result, \mode{\proj} not only scales efficiently with resolution but also further widens the gap in both performance and quality over prior approximation methods.

\subsection{Comparison with \mode{AdapTiV} and \mode{Exion}.}
\label{sec:eval:adaptiv_exion}

\hl{
\mbox{\Fig{fig:adaptiv_exion}} compares \mbox{\proj} with \mbox{\mode{AdapTiV}} and \mbox{\mode{Exion}} under different configurations.
The configurations are annotated in \mbox{\Fig{fig:adaptiv_exion}}.
Although \mbox{\mode{AdapTiV}} and \mbox{\mode{Exion}} achieve higher speedup with more aggressive token merging or sparsity, their generative quality degrades substantially as the compression ratio increases.
In contrast, \mbox{\proj} achieves higher speedup while maintaining much higher PSNR, demonstrating that channel-wise reuse better preserves video quality while still providing strong acceleration.
}
\section{Related Work}
\label{sec:realted}

\para{Diffusion Acceleration Techniques.}
\hl{
Algorithmically, early diffusion acceleration methods primarily focus on reducing the number of denoising timesteps, using techniques such as DDIM~\mbox{\cite{song2020denoising}}, DPM-Solver~\mbox{\cite{lu2022dpm1, lu2022dpm2}}, and flow matching~\mbox{\cite{lipman2022flow}}. 
Subsequent work further exploits similarity across timesteps to skip less important denoising steps, e.g., PAB~\mbox{\cite{zhao2024real}}, Astraea~\mbox{\cite{liu2025astraea}}, and ToCa~\mbox{\cite{zou2024accelerating}}. 
In contrast, \mbox{\proj} operates at the channel level, allowing different channel groups to be reused differently according to the spatio-temporal dimensions they encode. 
This finer granularity can preserve better visual fidelity.
Meanwhile, another line of research explores sparse patterns in attention scores~\mbox{\cite{jiang2024minference, xi2025sparse, yang2025sparsevideogen2, miao2026timeripple}} to reduce self-attention overhead. 
Instead, \mbox{\proj} reuses previously computed partial attention results for correlated channel values.
Thus, rather than discarding attention computations, \mbox{\proj} preserves them through reuse.
Diffusion caching methods~\mbox{\cite{liu2024timestep, chen2024delta}} exploit redundancy across denoising timesteps by reusing intermediate activations. 
\mbox{\proj} targets a different source of redundancy.
Therefore, \mbox{\proj} is complementary to diffusion caching methods.
}

\para{Diffusion Accelerators.}
\hl{
More recently, the hardware community has also focused on diffusion acceleration~\mbox{\cite{kong2024cambricon, you2023vitcod, heo2025exion, kim2025ditto, yoo2024adaptiv, zou2025s}}, proposing a range of accelerator designs that exploit computation sparsity and timestep similarity.
For example, AdapTiV~\mbox{\cite{yoo2024adaptiv}} merges similar tokens during execution to reduce inference computation, while EXION~\mbox{\cite{heo2025exion}} exploits the similarity of attention sparsity across denoising steps to reduce masking overhead.
Other works, such as Cambricon-D~\mbox{\cite{kong2024cambricon}} and Ditto~\mbox{\cite{kim2025ditto}}, leverage mixed precision to reduce computation overhead.
However, none of these studies explores the unique channel-wise spatio-temporal correlations in the video latent space.
In contrast, we are the first work that identifies the root cause of pattern diversity and proposes a principled reuse-based strategy to accelerate all vDiTs.
}

\para{Sparse Acceleration.}
A broad body of prior work has explored sparse computation techniques~\cite{zhu2019sparse, wang2021spatten, gonfimalla2019sparten, jang2021sparsity, sparch2020zhang, qin2020sigma, lu2021sanger, liu2022s2ta, pal2018outerspace, jang2019mnnfast, srivastava2020matraptor, wu2023highlight, ham2020a3, chen2019eyeriss}. 
Early efforts primarily focused on accelerating sparse matrix–matrix multiplication. 
E.g., OuterSPACE~\cite{pal2018outerspace}, SpArch~\cite{sparch2020zhang}, and MatRaptor~\cite{srivastava2020matraptor} explore different dataflows to improve computational efficiency, while SIGMA~\cite{qin2020sigma} and STC~\cite{zhu2019sparse} address irregular sparsity.
As DNNs became the dominant workload, subsequent work focused on exploiting sparsity in DNNs. 
For instance, Eyeriss~\cite{chen2019eyeriss} improves flexibility for compact and sparse models, while SparTen~\cite{gonfimalla2019sparten} and S2TA~\cite{liu2022s2ta} target different sparsity patterns. HighLight~\cite{wu2023highlight} further addresses workload imbalance in sparse computation.
More recently, research has focused on accelerating attention mechanisms. 
For example, A$^3$\cite{ham2020a3}, SpAtten~\cite{wang2021spatten}, and Sanger~\cite{lu2021sanger} reduce attention overhead through techniques such as approximation, token pruning, and quantization.
In contrast, our work targets a fundamentally different opportunity and exploits the channel-wise spatio-temporal correlations unique to vDiTs.

\para{Spatio-Temporal Similarity.}
There is a long history of exploiting spatio-temporal correlations to improve the efficiency of continuous vision~\cite{buckler2018eva2, zhu2018euphrates, feng2019asv, ying2022exploiting, song2020vr, zhao2020deja, zhao2021holoar, mahmoud2018diffy}. 
For example, both EVA$^2$\cite{buckler2018eva2} and Euphrates~\cite{zhu2018euphrates} observe the continuity in videos and propose motion-guided reuse techniques to reduce CNN inference cost. 
Subsequent work, e.g., Diffy~\cite{mahmoud2018diffy} and ASV~\cite{feng2019asv}, extends this idea to other vision tasks. 
VR-DANN~\cite{song2020vr} and CMC~\cite{song2024cmc} further exploit codec metadata to guide efficient video understanding. 
Meanwhile, Deja View~\cite{zhao2020deja} and Cicero~\cite{feng2024cicero} show that spatio-temporal correlations can also be used to improve the efficiency of VR applications.

\section{Conclusion}
\label{sec:conc}

As video generation continues to advance rapidly, the next frontier of generative AI will be vDiT models that can fundamentally understand the physical world. 
This paper introduced \proj, an algorithm–hardware co-design that rethinks how to accelerate video diffusion transformers. 
By exploiting channel-wise spatio-temporal correlations in the latent space, we reveal the fundamental relationships between the token channels and their attention scores.
By leveraging this insight, we proposed a lightweight reuse algorithm with a co-designed accelerator to efficiently address the irregular sparsity. 
Our principled design achieves up to 5.9$\times$ speedup and 16.0$\times$ energy savings over state-of-the-art GPUs and accelerators.
\begin{acks}

This work was supported by the Fundamental and Interdisciplinary Disciplines Breakthrough Plan of the Ministry of Education of China (JYB2025XDXM113), the National Natural Science Foundation of China (NSFC) Grants (62532006 and 62402312), Shanghai Pujiang Talent Program (24PJA044), and Shanghai Qi Zhi Institute Innovation Program (SQZ202316).

\end{acks}
% \clearpage

%%%%%%% -- PAPER CONTENT ENDS -- %%%%%%%%

\bibliographystyle{ACM-Reference-Format}
\bibliography{refs}

\end{document}